\documentclass{emulateapj}
\usepackage{psfig}
\usepackage{apjfonts,times}

\newcommand{\beq}{\begin{equation}}
\newcommand{\eeq}{\end{equation}}

\newcommand{\lapprox}{$\stackrel {<}{_{\sim}}$}
\def\alp{\mbox{$\alpha$}}

\def\farcm{\hbox{$.\mkern-4mu^\prime$}}
\def\farcs{\hbox{$.\!\!^{\prime\prime}$}}
\def\earth{\hbox{$\oplus$}}
\def\arcmin{\hbox{$^\prime$}}
\def\arcsec{\hbox{$^{\prime\prime}$}}
\def\solar{\mbox{$_{\normalsize\odot}$}}

\def\deg{\hbox{$^\circ$}}

\newcommand{\AmS}{{\protect\the\textfont2
  A\kern-.1667em\lower.5ex\hbox{M}\kern-.125emS}}
\newcommand{\lsim}{\ \raise
-2.truept\hbox{\rlap{\hbox{$\sim$}}\raise5.truept\hbox{$<$}\ }}
\newcommand{\gsim}{\ \raise
-2.truept\hbox{\rlap{\hbox{$\sim$}}\raise5.truept\hbox{$>$}\ }}
\newcommand{\simsim}{\ \raise
-2.truept\hbox{\rlap{\hbox{$\sim$}}\raise5.truept\hbox{$\sim$}\ }}

\slugcomment{Accepted for Publication in ApJ October 2, 2007}

\shorttitle{Star Formation and Clustering in NGC~346/N~66 in the SMC}
\shortauthors{E. Hennekemper et al.}

\begin{document}

\title{NGC~346 in the Small Magellanic Cloud. III. Recent Star Formation
and Stellar Clustering Properties in the Bright HII Region
N~66\altaffilmark{1}}


\author{Eva Hennekemper, Dimitrios A. Gouliermis, Thomas Henning,
Wolfgang Brandner}

\affil{Max-Planck-Institute for Astronomy, K\"onigstuhl 17, 69117
Heidelberg, Germany}

\and

\author{Andrew E. Dolphin}

\affil{Raytheon Corporation, USA}


\altaffiltext{1}{Research supported by the Deutsche Forschungsgemeinschaft (German
Research Foundation)}


\begin{abstract}

In the third part of our photometric study of the star-forming region
NGC~346/N~66 and its surrounding field in the Small Magellanic Cloud
(SMC), we focus on the large number of low-mass pre-main sequence (PMS)
stars revealed by the Hubble Space Telescope Observations with the
Advanced Camera for Surveys. We investigate the origin of the observed
broadening of the pre-main sequence population in the $V-I$, $V$ CMD.
The most likely explanations are either the presence of differential
reddening or an age spread among the young stars.  Assuming the latter,
simulations indicate that we cannot exclude the possibility that stars
in NGC 346 might have formed in two distinct events occurring about 10
and 5 Myr ago, respectively. We find that the PMS stars are not
homogeneously distributed across NGC 346, but instead are grouped in at
least five different clusters. On spatial scales from 0.8$''$ to 8$''$
(0.24 to 2.4\,pc at the distance of the SMC) the clustering of the PMS
stars as computed by a two-point angular correlation function is
self-similar with a power law slope $\gamma \approx -0.3$. The
clustering properties are quite similar to Milky Way star forming
regions like Orion OB or $\rho$\,Oph.  Thus molecular cloud
fragmentation in the SMC seems to proceed on the same spatial scales as
in the Milky Way. This is remarkable given the differences in
metallicity and hence dust content between SMC and Milky Way star
forming regions.

%

\end{abstract}


\begin{figure*}[t!]
\epsscale{1.15}
\plottwo{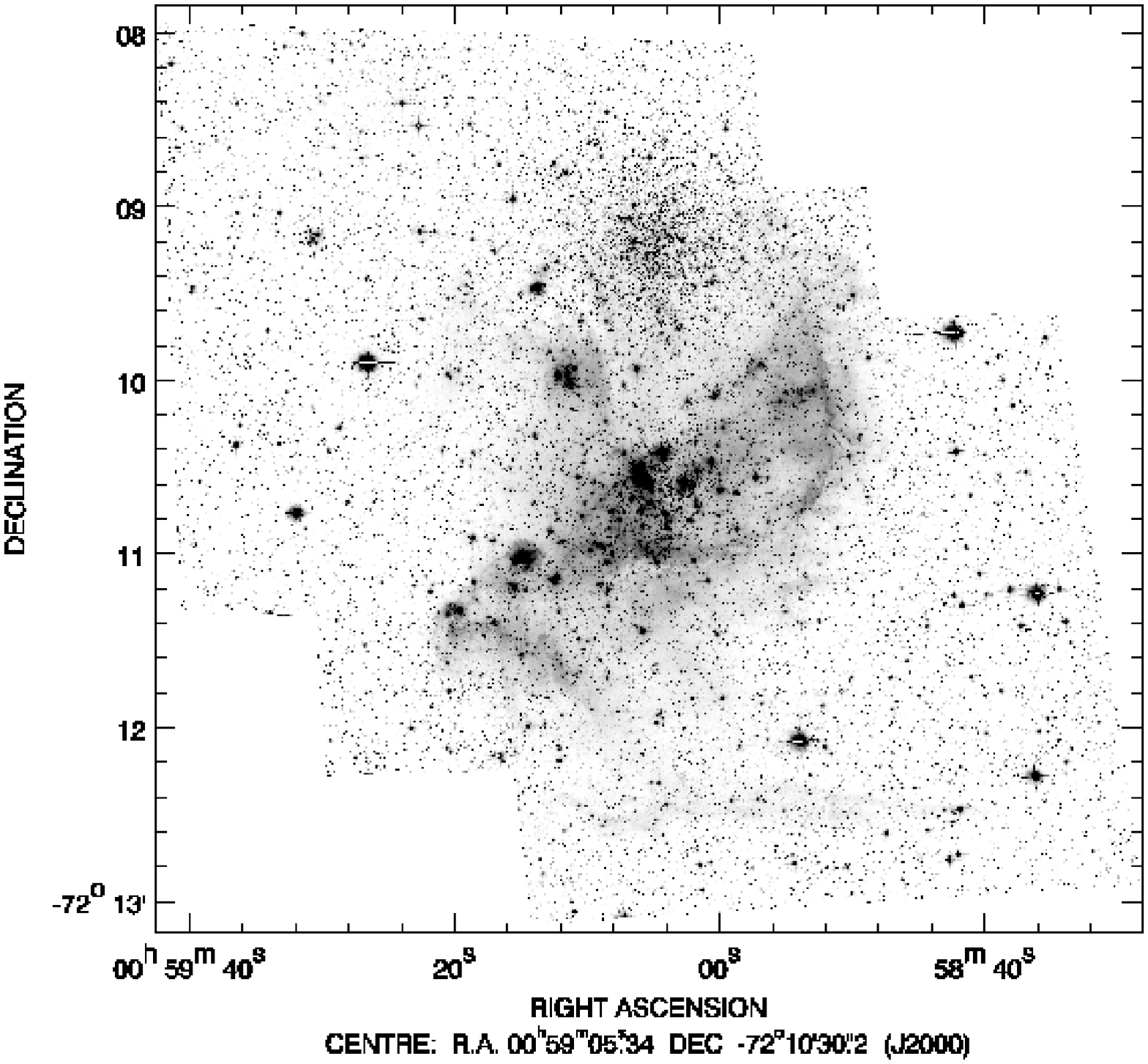}{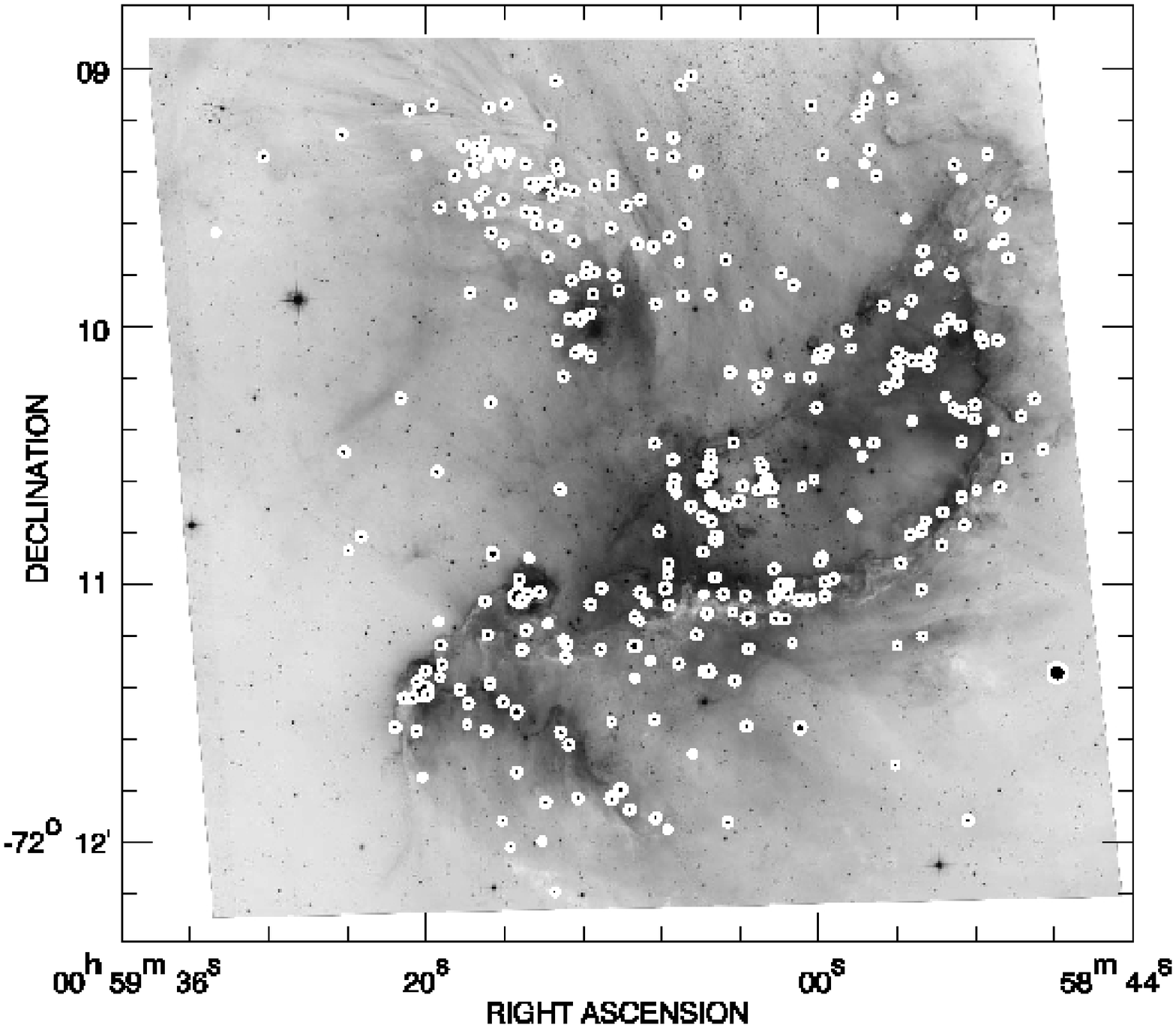}
\caption{Left: ACS/WFC observations of NGC 346,
composed of three overlapping pointings in the $I$-band.
The association NGC~346 covers the
center of the field, located almost in the middle of the ``bar" of N~66,
as it is outlined by the nebulosity. The intermediate-age cluster BS~90
(Rochau et al. 2007) is seen as the circular stellar concentration to
the north of the association.
{\em Right:} On the ACS/WFC H\alp\ image of NGC~346/N~66 
stars with intrinsic H\alp\ excess are marked by circles.
The spatial distribution of the excess stars outlines the
H{\sc ii} region, and coincides with the overall distribution
of the PMS stars.}
\label{fig-ima}
\end{figure*}

\section{Introduction}

It is well known that Galactic OB associations also host large numbers
of fainter, low-mass pre-main sequence (PMS) stars (e.g.\ Preibisch \&
Zinnecker 1999; Sherry et al.\ 2004; Brice\~{n}o et al.\ 2005). 
Low-mass PMS stars in stellar associations provide a longer-lasting
record of the most recent star formation events than the short-lived
high-mass stars.  Large-scale surveys have identified hundreds of PMS
stars in nearby OB associations (Brice\~{n}o et al. 2007). In order to
understand the star formation triggering and history, or the Initial
Mass Function (IMF), one has to study both high- and low-mass stars in
star forming regions.

In many cases the low-mass populations of galactic OB associations
cannot be easily distinguished from foreground main-sequence stars (e.g.
Brice\~{n}o et al. 2001).  Recently PMS stars have been discovered for
the first time in an extragalactic stellar association, which suffers
much less from contamination by the galactic disk.  {\em Hubble Space
Telescope} (HST) observations of the association LH\,52 in the Large
Magellanic Cloud (LMC) revealed $\approx$500 PMS stars with masses down
to 0.3\,M$_\odot$ (Gouliermis et al. 2006a). The investigation of
individual extragalactic pre-main sequence stars opens a new field of
study. As both high angular resolution and wide field-of-view are
required, HST is the ideal observatory to carry out these studies. While
HST observations of extragalactic associations cannot reach the
detection limits achieved for local associations, they provide a unique
opportunity for studying low-mass star formation in other galaxies. 

The OB association NGC~346 in the Small Magellanic Cloud (SMC) is
located in the central part of the brightest {\sc H ii} region in the
SMC, named LHA~115-N~66 or in short N~66 (Henize 1956). With 33
spectroscopically confirmed O and B stars, NGC~346 hosts the largest
sample of young, massive stars in the SMC (Walborn 1978; Walborn \&
Blades 1986; Niemela et al. 1986; Massey et al. 1989; Walborn et al.
2000; Evans et al.  2006).  Photoionization models by Rela\~{n}o et al.
(2002) imply that these stars are a major source of ionizing flux for
the surrounding diffuse interstellar medium (ISM). Recent imaging from
the Wide-Field Channel (WFC) of the {\em Advanced Camera for Surveys}
(ACS) on-board HST (GO Program 10248; PI: A. Nota) revealed the PMS
stellar content of the general region of NGC~346/N~66 down to the
sub-solar mass regime (Nota et al.\ 2006; Gouliermis et al.\ 2006b --
hereafter Paper I -- ; Sabbi et al.\ 2007).  Nota et al.\ (2006) suggest
that all PMS stars in the association are the product of a single star
formation event, taking place 3 to 5 Myr ago.  The PMS distribution in
the $V-I$, $V$ color-magnitude diagram (CMD), however, shows a prominent
widening, which may be explained by an age spread of $\approx$10 Myr.
This raises the question {\em if the PMS stars in NGC\,346 are indeed
the result of a single star formation event.}

Sabbi et al.\ suggest that the PMS population is mainly concentrated in
a number of subclusters (three of them at the central part, where the
association is located), which formed at the same time from the
turbulence-driven density variations, and not following a sequential
process.  Star counts, however, reveal only a few compact PMS clusters
with a significant number of stars in the area around the association.
The main body of the association cannot be divided into separate
subclusters (Paper I). {\em Could the remote clusters be the product of
a star formation event occurring before or after the event which
triggered the formation of the association?}

In the first part of our study of this extraordinary star-forming region
(Paper I), we compiled our ACS photometry of almost 100,000 stars, and
presented preliminary results on stellar types and their distinct
spatial distributions. We confirmed the co-existence of massive OB stars
and low-mass PMS stars in NGC 346. In the present paper, we explore in
greater detail the properties of the low-mass PMS stars with a focus on
clustered star formation in NGC~346/N~66.

The paper is organized as follows: In \S~2 we describe the datasets from
HST/ACS, present the color-magnitude diagram, carry out a comparison of
our photometry of the brightest stars with previous photometric studies,
and discuss the reddening in the region. The broadening of the PMS
distribution in the CMD, a thorough discussion of possible causes and
explanations of this phenomenon as well as the spatial variations in the
CMD are presented in \S~3.  In \S~4 we present the clustering properties
of the PMS stars and discuss them in terms of hierarchical star
formation. Our analysis of H\alp\ observations with HST/ACS as well as
previous {\em Spitzer} Space Telescope results on Young Stellar Objects
(YSOs) are presented in \S~5 and \S~6, respectively. In \S~7 we
summarize the results.

\section{Description of the Data\label{sec-dat}}

The observations used in this study, taken with HST/ACS (GO Program
10248), and the photometry with the package {\tt DOLPHOT}\footnote{The
ACS mode of {\tt DOLPHOT} is an adaptation of the photometry package
{\tt HSTphot} (Dolphin 2000). It can be downloaded from {\tt
http://purcell.as.arizona.edu/dolphot/}.} is thoroughly presented in
Paper I. These observations cover an area of about 5\arcmin\ $\times$
5\arcmin\ wide from imaging of three overlapping ACS fields observed
with the Wide-Field Channel (WFC) of the camera centered on NGC~346
(Table 1 of Paper I) in the filters $F555W$ and $F814W$ ($\equiv V, I$).
The $F814W$ image of the whole field is shown in Fig.~\ref{fig-ima}
(left). This area covers the intermediate-age cluster BS 90 (Bica \&
Schmidt 1995), the association NGC~346, and all known components of the
bright nebula N~66, also known as DEM L 103 (Davies, Elliott \& Meaburn
1976). The field covering the central area, at RA (J2000) = 00$^{\rm
h}$59$^{\rm m}$07$^{\rm s}$, DEC (J2000) =
$-$72{\deg}10{\arcmin}31{\arcsec}, was also observed in the narrow-band
filter $F658N$, equivalent to H\alp\ (``Advanced Camera for Surveys
Instrument Handbook'' Ver. 7, Oct 2006) with three 512s exposures
(Fig.~\ref{fig-ima}, {\em right}). In the present study we used {\tt
DOLPHOT} to obtain photometry from the H\alp\ images, and to combine the
results with our $V$ and $I$ photometry (see \S~5). 

\begin{figure*}[]
\epsscale{1.15}
\plottwo{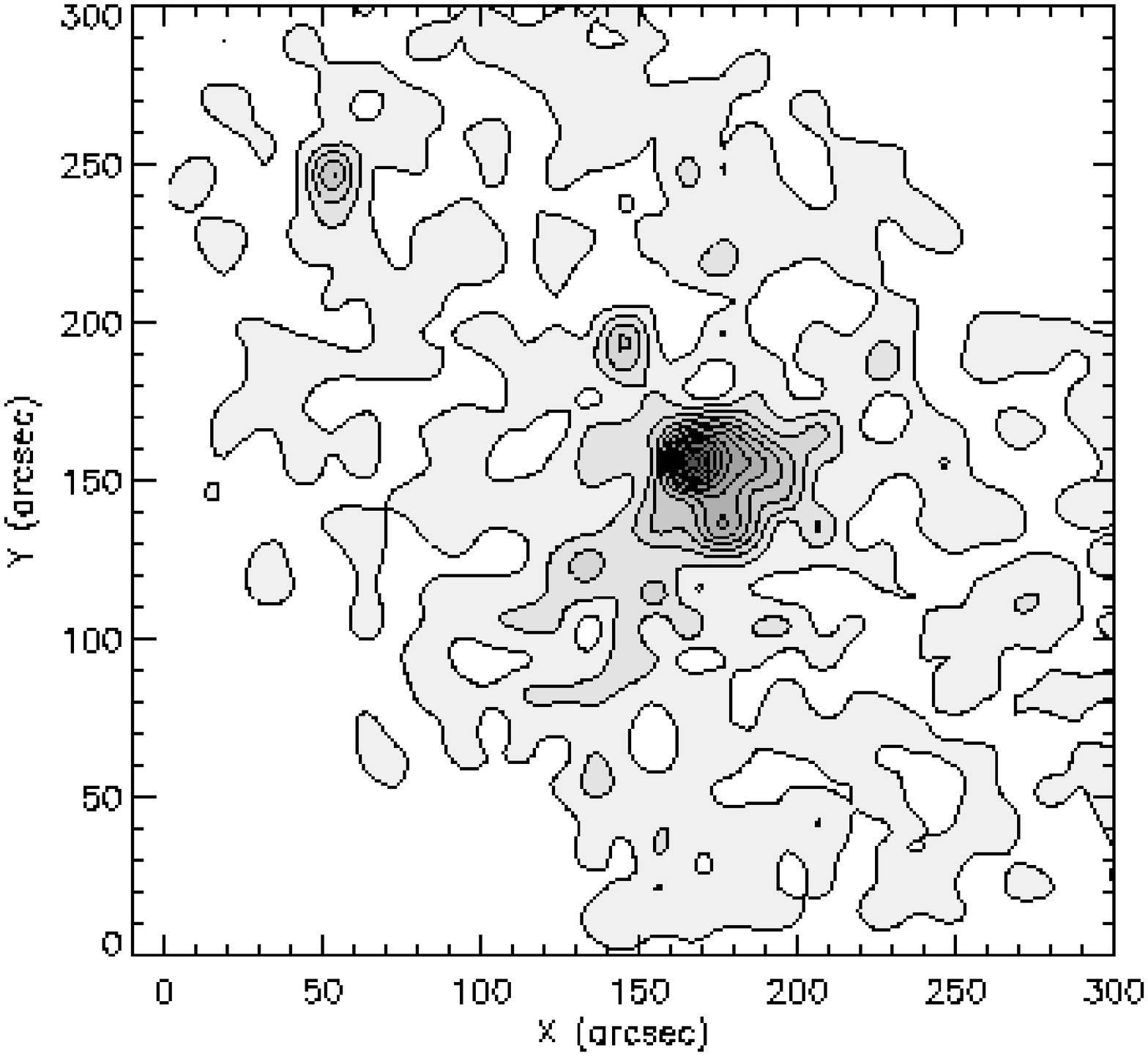}{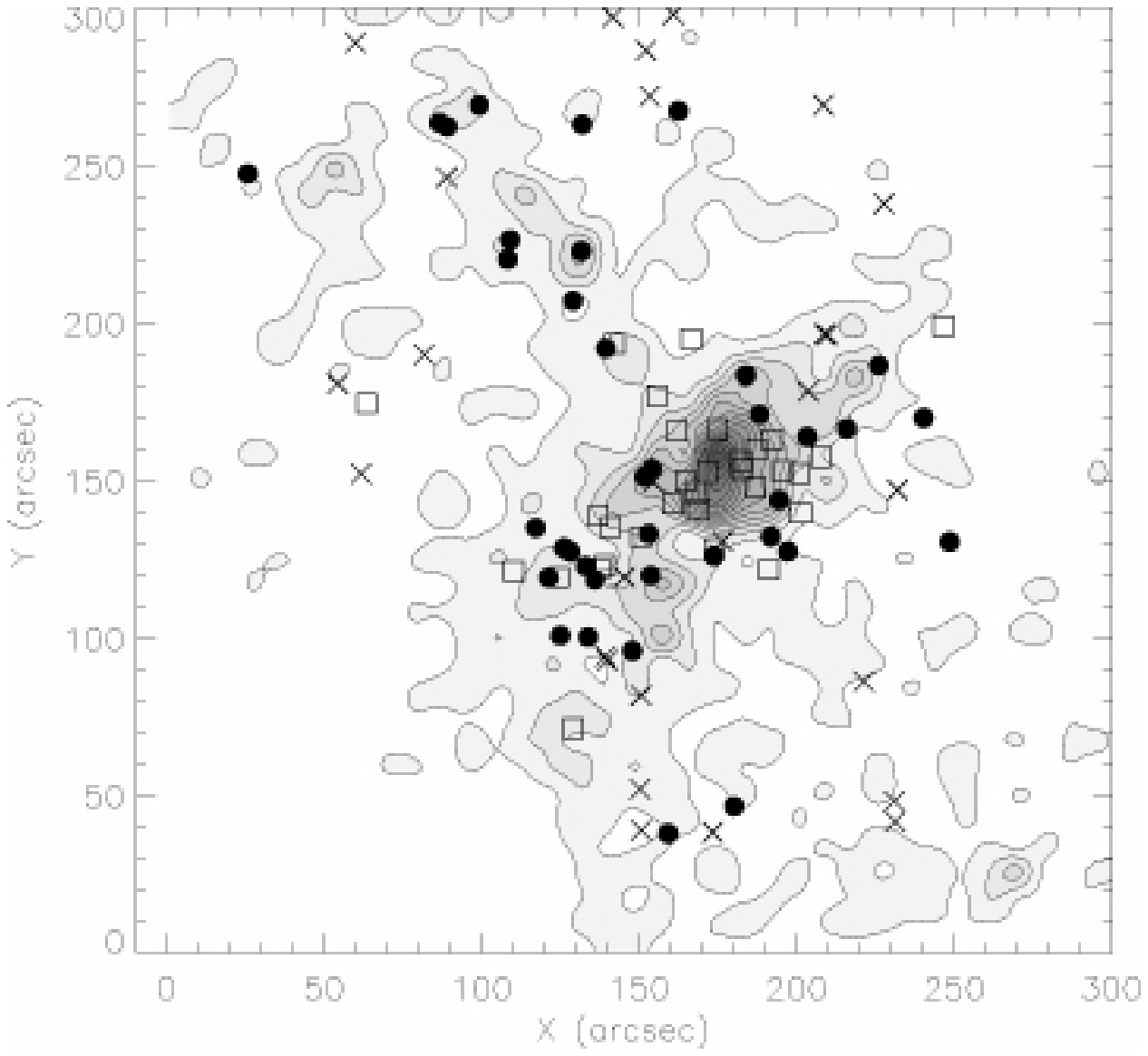}
\caption{Isodensity contour maps constructed from star counts of Upper 
Main Sequence ({\em left}), and Pre-Main Sequence ({\em right}) stars, as 
they have been selected in Paper I according to their distinct positions 
in the CMD. The lowest isopleth corresponds to the background surface 
density and isopleths of higher surface density are plotted in steps of 
1$\sigma$, where $\sigma$ is the standard deviation of the background 
density. On the right, the positions of candidate YSOs found in the 
S$^3$MC Spitzer Survey in the observed area around NGC~346 (Simon et al. 
2007) are also plotted with crosses and dots, while the brightest 31 OB 
stars, spectroscopically verified by Massey et al. (1989), included in our 
photometry are plotted with open boxes. North is up and east to the left 
on the map.\label{fig-cms}}
\end{figure*}

\subsection{The Color-Magnitude Diagram of NGC~346}

Based on star counts performed in Paper I, we identified the areas where
stars of different types are concentrated. The isodensity contour maps
from counting the bright upper main sequence (UMS) and PMS stars are
shown in Fig.~\ref{fig-cms}. In Paper I these maps allowed the
identification of the region which is covered by the association alone.
This region is confined within a radius of about 0\farcm6 around the
geometric center of the system. The $V-I$, $V$ Color-Magnitude Diagram
(CMD) of the stars detected in this region, as the most representative
of the association NGC~346, is shown in Fig.  \ref{fig-cmdsys}. This CMD
is characterized by a very bright upper main sequence and a prominent
sequence of PMS stars, both contaminated by the stellar content of the
general field of the SMC, but also of the second dominant stellar
concentration, the cluster BS 90. We also identified the area covered by
the latter, which is the subject of the second part of our study (Rochau
et al.  2007). The part we present here deals with the recent star
formation of this active region, and thus it focuses on the young
populations as they are revealed from the bright UMS and the faint PMS
stars found by our photometry.

\begin{deluxetable*}{rcccccclcccc}
\tablewidth{0pt}
\tabletypesize{\scriptsize}
\tablecaption{Photometry of OB stars in NGC 346 by Massey et al. (1989),
and from our ACS analysis. Colors and spectral types are also taken from
Massey et al. (1989). The corresponding reddening free parameters $Q$
and the derived individual un-reddened colors, color excesses and
extinctions are also given. Star ID numbers and celestial coordinates
(in J2000) are from our catalog published in Paper I. Units of
declination are degrees, arcminutes, and arcseconds; units of right
ascension are hours, minutes and seconds.\label{tab1}}
\tablehead{
\colhead{No} & 
\colhead{RA} & 
\colhead{DEC} & 
\colhead{$V_{\rm ACS}$} & 
\colhead{$V_{\rm CTIO}$} & 
\colhead{($U-B$)$_{\rm CTIO}$} & 
\colhead{($B-V$)$_{\rm CTIO}$} & 
\colhead{Sp. Type} & 
\colhead{$Q$} & 
\colhead{($B-V$)$_{\rm 0}$} & 
\colhead{$E(B-V)$} & 
\colhead{$A_{\rm V}$}
}
\startdata
 2&00:59:04.46 &$-$72:10:24.49&12.44&12.57&$-$1.09&$-$0.20&O5.5If     &$-$0.95&$-$0.31&0.11&0.37\\
 5&00:59:00.72 &$-$72:10:27.91&13.52&13.50&$-$1.12&$-$0.23&O3V((f*))  &$-$0.95&$-$0.32&0.09&0.28\\
 6&00:59:00.01 &$-$72:10:37.67&13.69&13.66&$-$1.05&$-$0.23&O5.5V      &$-$0.88&$-$0.29&0.06&0.20\\
 9&00:58:57.36 &$-$72:10:33.38&13.98&14.02&$-$1.05&$-$0.24&O4V((f))   &$-$0.88&$-$0.29&0.05&0.16\\
10&00:59:01.77 &$-$72:10:30.94&14.38&14.18&$-$1.08&$-$0.23&O5.5V((f$+$))&$-$0.91&$-$0.30&0.07&0.24\\
13&00:59:06.71 &$-$72:10:41.02&14.65&14.53&$-$1.01&$-$0.22&O6.5V      &$-$0.85&$-$0.28&0.06&0.20\\
14&00:59:06.31 &$-$72:09:55.84&14.37&14.26&$-$0.91&$-$0.10&B0.5V      &$-$0.84&$-$0.28&0.18&0.57\\
15&00:59:02.88 &$-$72:10:34.68&14.45&14.39&$-$1.08&$-$0.22&O7V	      &$-$0.92&$-$0.31&0.09&0.28\\
18&00:59:14.50 &$-$72:11:59.53&14.97&14.98&$-$0.90&$-$0.16&B0V	      &$-$0.78&$-$0.26&0.10&0.32\\
19&00:59:28.75 &$-$72:10:16.28&15.09&15.14&$-$0.95&$-$0.08&B0V	      &$-$0.89&$-$0.30&0.22&0.69\\
20&00:59:12.28 &$-$72:11:07.66&14.79&14.68&$-$0.96&$-$0.19&O7V	      &$-$0.82&$-$0.27&0.08&0.27\\
21&00:59:12.66 &$-$72:11:08.84&14.85&14.52&$-$0.96&$-$0.17&O8V	      &$-$0.84&$-$0.28&0.11&0.35\\
24&00:59:18.58 &$-$72:11:09.64&14.96&14.87&$-$1.02&$-$0.24&O8V	      &$-$0.85&$-$0.28&0.04&0.13\\
26&00:59:07.29 &$-$72:10:25.10&15.09&15.13&$-$0.95&$-$0.30&O8V	      &$-$0.73&$-$0.24&0.00&0.00\\
27&00:59:05.86 &$-$72:10:50.09&15.01&15.10&$-$1.00&$-$0.31&O8V	      &$-$0.78&$-$0.26&0.00&0.00\\
29&00:59:01.86 &$-$72:10:43.07&15.03&15.03&$-$1.00&$-$0.28&O9.5V      &$-$0.80&$-$0.27&0.00&0.00\\
30&00:59:12.78 &$-$72:10:52.14&15.18&15.00&$-$0.93&$-$0.21&O9.5V      &$-$0.78&$-$0.26&0.05&0.16\\
31&00:59:15.48 &$-$72:11:11.44&14.98&14.82&$-$0.98&$-$0.15&O6V	      &$-$0.87&$-$0.29&0.14&0.45\\
32&00:59:09.80 &$-$72:10:58.76&15.25&15.26&$-$0.96&$-$0.22&O8V	      &$-$0.80&$-$0.27&0.05&0.15\\
33&00:59:05.97 &$-$72:10:44.69&15.20&15.04&$-$0.96&$-$0.21&B0V	      &$-$0.81&$-$0.27&0.06&0.19\\
37&00:59:11.61 &$-$72:09:57.31&15.23&14.96&$-$1.01&$-$0.16&O5.5V      &$-$0.89&$-$0.30&0.14&0.44\\
38&00:58:58.74 &$-$72:10:51.10&15.21&15.20&$-$0.97&$-$0.27&O7.5V      &$-$0.78&$-$0.26&0.00&0.00\\
39&00:58:58.84 &$-$72:10:38.57&15.20&15.15&$-$0.97&$-$0.17&B0V	      &$-$0.85&$-$0.28&0.11&0.36\cr
42&00:59:05.17 &$-$72:10:38.24&15.20&15.20&$-$1.04&$-$0.18&B0V	      &$-$0.91&$-$0.30&0.12&0.39\cr
44&00:59:07.59 &$-$72:10:48.07&15.41&15.33&$-$0.95&$-$0.24&O6V	      &$-$0.78&$-$0.26&0.02&0.06\cr
48&00:59:04.76 &$-$72:11:02.69&15.39&15.33&$-$0.87&$-$0.18&O7.5V      &$-$0.74&$-$0.25&0.07&0.21\cr
53&00:59:11.88 &$-$72:10:55.56&15.68&15.62&$-$0.99&$-$0.20&B0V	      &$-$0.85&$-$0.28&0.08&0.26\cr
54&00:59:00.92 &$-$72:11:09.02&15.60&15.50&$-$1.03&$-$0.26&O6.5V      &$-$0.84&$-$0.28&0.02&0.06\cr
57&00:59:08.65 &$-$72:10:13.87&15.62&15.50&$-$1.02&$-$0.28&O7V	      &$-$0.82&$-$0.27&0.00&0.00\cr
65&00:59:02.82 &$-$72:10:37.20&15.63&15.53&$-$1.02&$-$0.22&B0V	      &$-$0.86&$-$0.29&0.07&0.21\cr
74&00:58:48.90 &$-$72:09:51.84&15.86&15.85&$-$0.91&$-$0.16&B0	      &$-$0.79&$-$0.26&0.10&0.33
\enddata
\end{deluxetable*}

\subsection{Completeness}

\begin{figure}[]
\epsscale{1.15}
\plotone{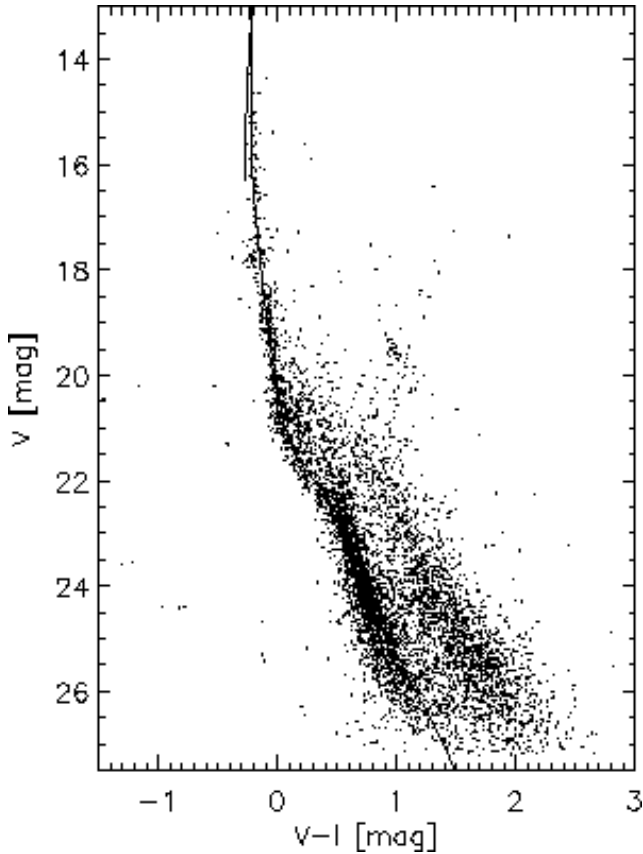}
\caption{The $V-I$, $V$ Color-Magnitude Diagram (CMD) of the association
NGC~346 from HST/ACS photometry. A region within a radius of
0\farcm6 is selected as the best representation of the stellar
population, which is characterized by two features in the CMD:
the upper main sequence, with 33 spectroscopically confirmed OB stars,
and the lower mass Pre-Main Sequence. An isochrone model for an age
of $\approx$4\,Myr Girardi et al.\ (2002) for  $E(B-V)=0.08$ mag and a
distance modulus $m-M=$ 18.9 mag is overplotted, and confirms
the young age of the association.}
\label{fig-cmdsys}
\end{figure}

\begin{figure}[t!]
\epsscale{1.15}
\plotone{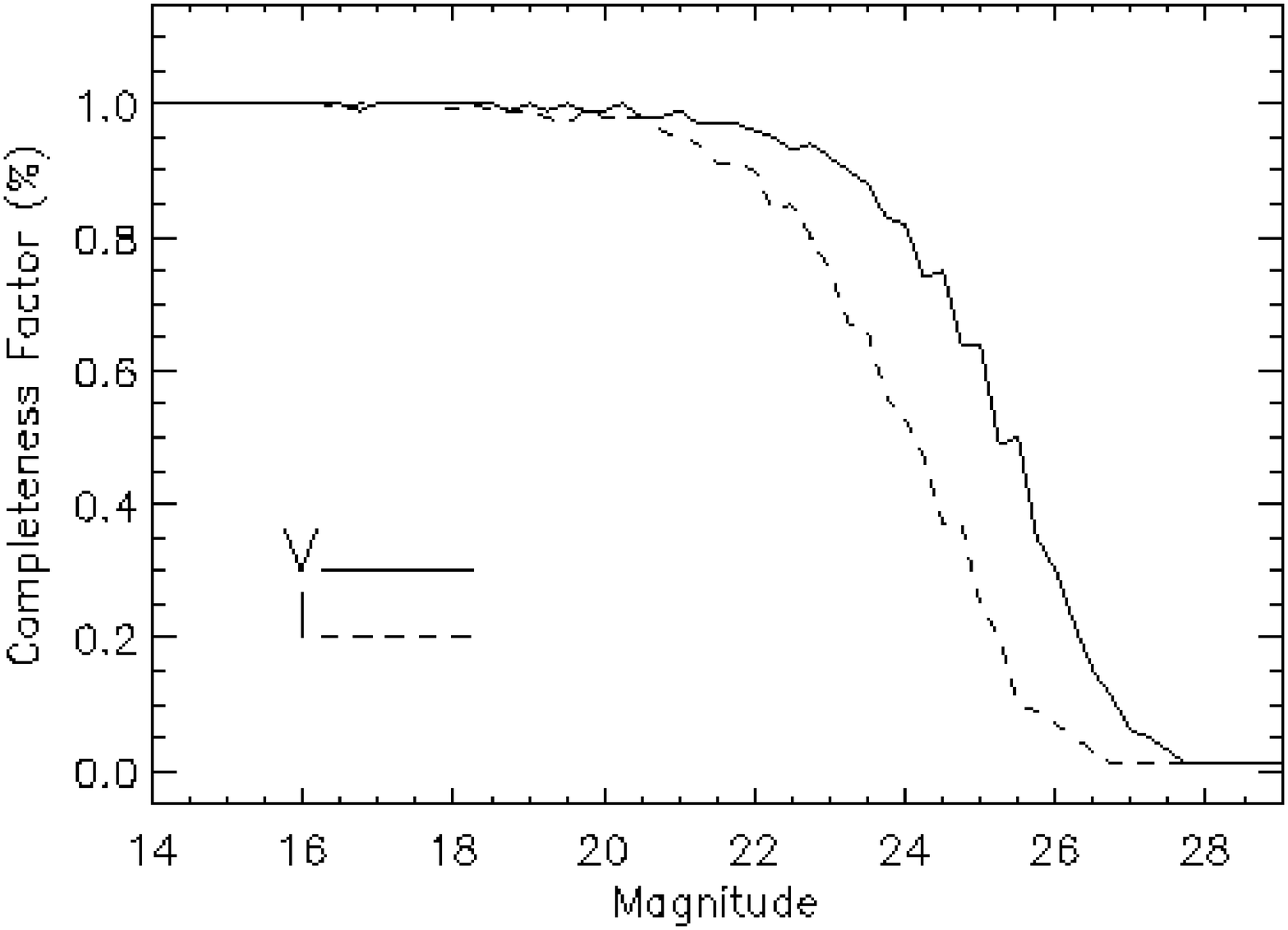}
\caption{Completeness factors as derived by {\tt DOLPHOT} for all V
and I band datasets in NGC~346 within a radius of about 0\farcm6 around 
its center. Note that due to crowding and the presence of diffuse emission,
the completeness is almost one magnitude worse than when considering the
entire stellar sample as is shown in Fig.\ 2 in Paper I.}
\label{fig-cmp}
\end{figure}

The photometric completeness is defined as a function of magnitude for
both $V$ and $I$ band by artificial star experiments described in Paper
I. In Fig. 2 of Paper I the ``overall'' completeness functions for the
whole observed stellar sample are shown. Here, we show the completeness
factors for the region of the association (Fig.~\ref{fig-cmp}), which
corresponds to the CMD of Fig.~\ref{fig-cmdsys}. A comparison of the
completeness in the area of the association with the overall
completeness of the entire region (Paper I) and the one in the area of
the cluster BS 90 (Rochau et al. 2007) shows that the association NGC
346 has the lowest completeness. This is not due to higher crowding,
since the area of BS 90 is much more crowded, but most probably due to
the higher extinction related to this stellar system.

\subsection{Comparison with previous catalogs \label{massey-compare}}

\begin{figure}[t!]
\epsscale{1.15}
\plotone{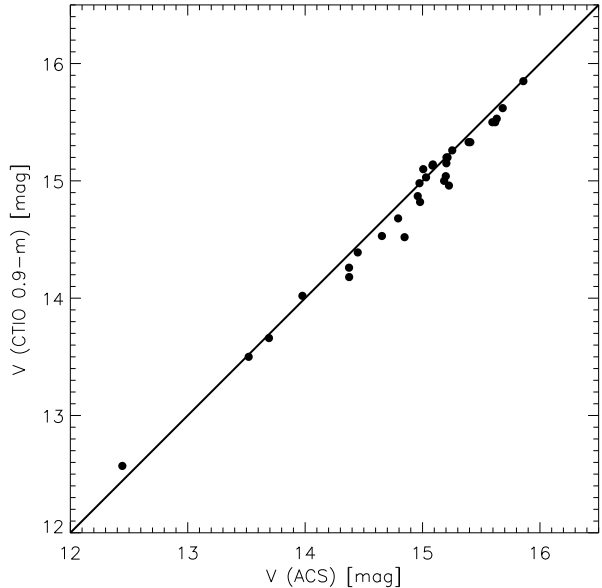}
\caption{Comparison between the $V$ magnitudes from our photometry with 
ACS and the photometry with the 0.9-m CTIO Telescope for the 31 brightest 
OB stars in NGC~346, for which Massey et al.  determined accurate spectral 
types. The drawn line corresponds to the line of excellent coincidence.}
\label{fig-phtcmp}
\end{figure}

Our photometry also includes the brightest stars in NGC~346. Since the 
bright main-sequence stars in NGC~346 have been photometrically and 
spectroscopically studied in detail by Massey et al. (1989), it is worth 
comparing our results with the ones from these authors. We found, thus, 
for each star in the Massey et al. (1989) catalog its corresponding record 
in our catalog. We identified 31 out of a total of 33 bright OB stars for 
which Massey et al. performed spectroscopy. Our photometry is in excellent 
agreement with the photometry of these authors, in the sense that their 
differences are negligible. This is demonstrated in Fig.~\ref{fig-phtcmp}, 
where the $V$ magnitudes from CTIO 0.9-m Telescope photometry by Massey et 
al. and our $V$ magnitudes from ACS for the 31 common stars are compared 
(see also Table \ref{tab1}).

\subsection{Reddening in the observed area\label{sec-red}}

In order to evaluate the reddening toward different regions of the
observed area, we estimated the reddening of individual stars by
comparing to ``un-reddened'' isochrone models. Specifically, the
``youngest'' available Padova isochrone in the ACS filter system
(Girardi et al. 2002) was compared with main sequence stars brighter
than $V\simeq 22$~mag (``upper main sequence'' population in Paper I).
The observed position in the CMD of each main sequence star with
$V$~\lsim~22~mag was compared to the expected position the star would
have according to the Padova isochrone model of $\log{\tau}=6.6$,
assuming that this is the age of the star. Metallicity of $Z=0.004$ and
a distance modulus of $m-M=18.9$~mag, both typical values for the SMC,
were considered for the model. The difference between the observed
magnitude and color of each star and the values expected by the model
provide a measurement of the color excess, $E(V-I)$, and extinction,
$A_{\rm V}$, for the star. The produced reddening map and the
corresponding reddening distribution are shown in Fig.~\ref{fig-genred}. 
This distribution peaks at $E(B-V) \simeq$~0.08~mag. The reddening seems
to vary from one region to the next and within the boundaries of the
association NGC~346 itself, as we show in the next section.


\subsection{Extinction around the OB stars of NGC~346\label{sec-difred}}

Differential reddening is important for the studies of young stellar
populations in bright massive stellar systems like NGC~346. In order to
quantify the interstellar reddening in different regions within the
association, we make use of previous spectroscopic studies of the
brightest main-sequence stars observed with our photometry. We have
already matched these stars with the OB stars of NGC~346 studied by
Massey et al. (1989) (see \S \ref{massey-compare}). Therefore, we know
their exact spectral types. In addition, the $UBV$ photometry of these
authors can be used for estimating the reddening-free parameter $Q$:
$$Q=(U-B)-0.72 \times (B-V)$$ Based on the spectral classification of OB
stars in our Galaxy by Schmidt-Kaler (1982) and Peletier (1989) and the
photometry of standards of such spectral types by Johnson \& Morgan
(1953), it is known that the ``un-reddened'' color index $(B-V)_{\rm 0}$
is related to $Q$ by a simple function (Binney \& Merrifield 1998): $$
(B-V)_{\rm 0} \simeq 0.332 Q$$ We used this relation to estimate
$(B-V)_{\rm 0}$ for each of our OB stars with known spectral types from
Massey et al. (1989). Therefore, we are able to estimate their
individual color excess $E(B-V)$ by comparing the observed color indices
$(B-V)$ to their un-reddened ones $(B-V)_{\rm 0}$, as calculated from
their $Q$ parameters, and consequently their individual extinction
$A_{\rm V}$. The extinction and reddening values found for each OB star
are given in Table \ref{tab1}, where it is shown that individual
reddening for the OB stars varies from $E(B-V) \simeq$ 0.0 to 0.2 mag
(0.0 \lsim\ $A_{\rm V}$ \lsim\ 0.7 mag). 

\begin{figure*}[]
\epsscale{1.15}
\plottwo{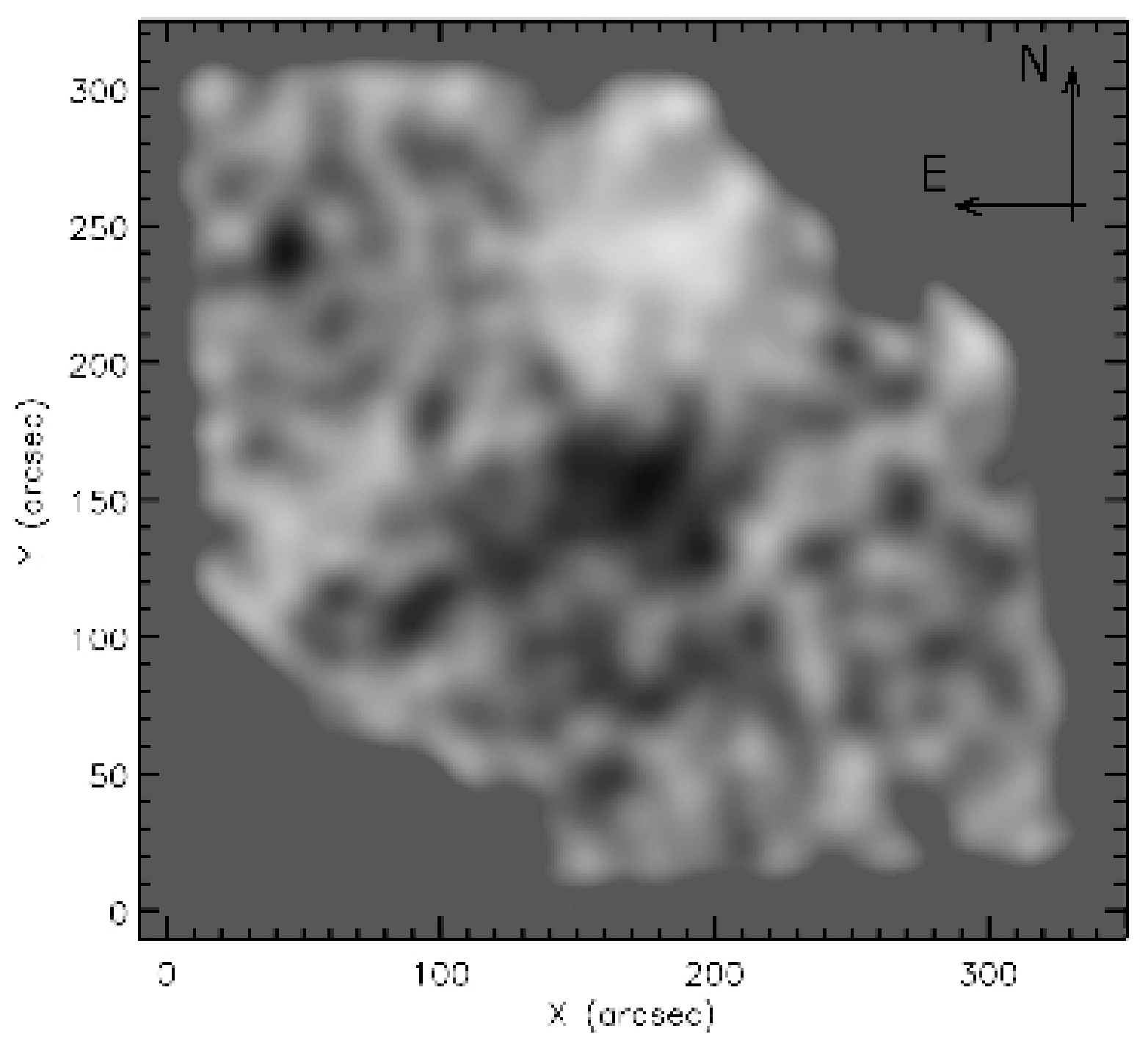}{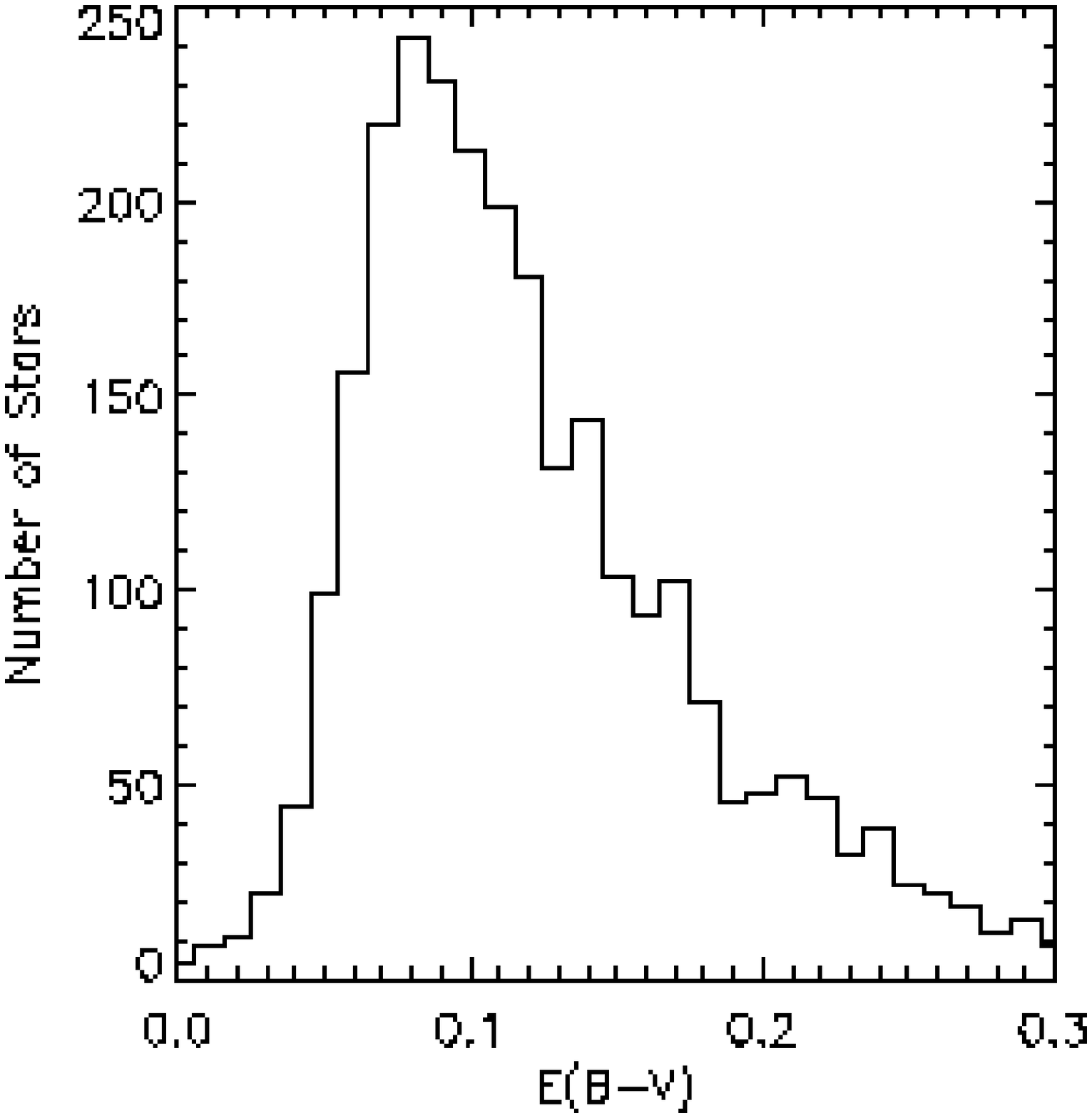}
\caption{{\em Left}: Reddening Map around NGC~346. This map is based on the 
upper main sequence stars
down to $V = 22.15$~mag, as they were selected in Paper I. The reddening
for each star was estimated by comparison to a
Padova isochrone for an age of $\approx$4\,Myr (Girardi et al. 2002). Darker
regions, indicating high reddening, are apparent in the general region of
the association and along the related molecular filaments of the N~66
``bar''. The white area to the north corresponds to the cluster BS~90,
where no upper main sequence stars are found (Rochau et al. 2007), and
therefore there is no reddening measurement for this area. This map also
demonstrates the patchy behavior of the reddening further away from the
association. {\em Right}: The reddening distribution from the upper main
sequence stars, as it is estimated from comparison with isochrone
models. The peak corresponds to color excess $E(B-V)\simeq0.08$ mag.}
\label{fig-genred} 
\end{figure*}

The OB stars are centrally concentrated in the association NGC~346
itself, as shown from their spatial distribution in Fig.~\ref{fig-cms}
(right; their positions are indicated by the open boxes), and therefore
a comparison between the spatial distribution of their reddening and the
one derived from all upper main sequence stars for the whole observed
region is not possible. However, typical reddening values derived for
the OB stars in the area of the association are found to be
systematically lower than those derived in the previous section from
upper main sequence stars in the same location. Moreover, the reddening
found for the OB stars follow a normal number distribution peaked at
$E(B-V) \simeq$~0.04~mag, which is half of the one found from the upper
main sequence stars of the whole region (shown in
Fig.~\ref{fig-genred}). From the spatial distribution of the reddening
derived from the OB stars, we find that $A_{\rm V}$ is higher in the
inner parts of the association with values between $A_{\rm V} \simeq$
0.2 and 0.4 mag ($E(B-V) \simeq$ 0.06 - 0.12 mag). A possible
explanation for the lower reddening toward the OB stars is that probably
strong winds from these stars have blown away the gas around them, and
therefore reddening appears lower in their vicinities. Considering that
the reddening estimation in the previous section is based on a larger
statistical sample of stars than the one presented here only for the OB
stars, in the subsequent analysis we assume a mean interstellar
reddening of $E(B-V) \simeq$~0.08~mag, a result which also agrees with
the one derived for the same region by Sabbi et al. (2007).

\section{Pre-Main Sequence Stars in NGC~346 \label{sec-pms}}

The ages of low-mass PMS stars are usually determined from their
location in the CMD by comparison to theoretical isochrones (e.g.
Brice\~{n}o et al.  2007). Quite often, a widening of the observed PMS
in the CMD gives evidence for an age spread (Palla \& Stahler, 2000).
However, this spread could also occur because of several other factors
which can cause considerable deviations of an individual star's position
in the CMD from the location predicted by theoretical models. Such
factors include variability, binarity, and the distribution of distances
along the line of sight. According to Shu's et al. (1987) picture of
star formation a typical galactic molecular cloud could sustain star
formation for a timescale on the order of the diffusion time,
$\sim10^{7}$~yrs, and an observable consequence would be an age spread
in star forming regions, with age of PMS stars comparable to this
timescale.  Still, as Brice\~{n}o et al. (2007) reports, no evidence for
a wide age spread has been found in the PMS populations of $\sigma$~Ori,
Ori OB1, $\lambda$~Ori and Upper Sco in our Galaxy. These authors
conclude that the data from these different star forming regions support
the hypothesis that star formation occurs fast and synchronized in
molecular clouds. {\em Do we observe the same behavior in star-forming
regions of the MCs like NGC~346/N~66 based on our ACS photometry of its
PMS population?} In the following paragraphs we explore this question.

\subsection{Location of PMS Stars in the CMD \label{sec-pmsloc}}

\begin{figure*}[]
\epsscale{1.}
\plotone{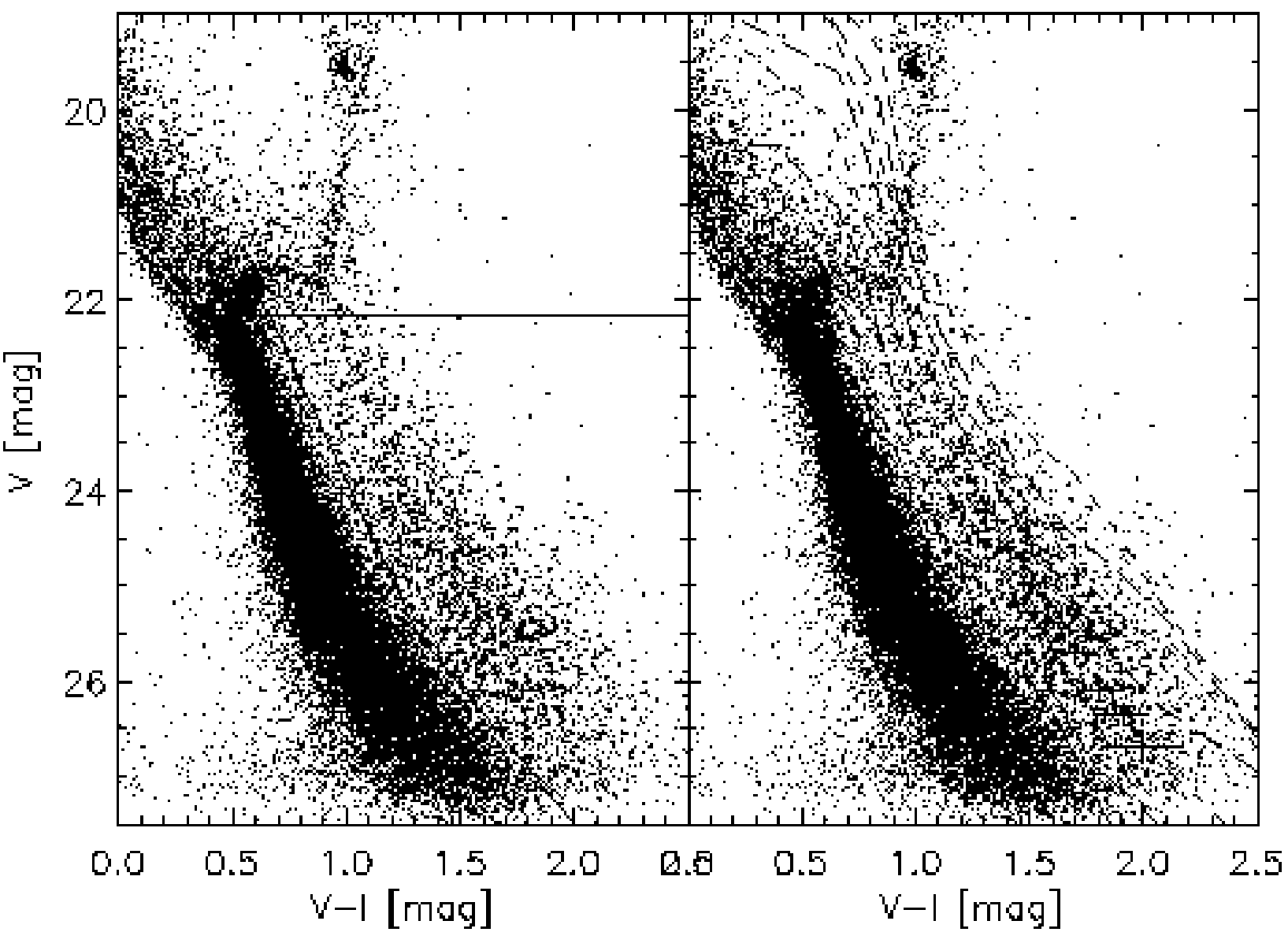}
\caption{$V-I$, $V$ Color-Magnitude Diagram (CMD) of all
stars detected in the area of NGC~346 with ACS imaging, centered on the
region of the observed PMS stars. The plot in the left panel shows the
area selected in Paper I as the best representative of these
stars. In the right panel PMS isochrones by Siess et al. (2000) for ages 
of 0.5, 0.75, 1.0, 1.5, 2.5, 5, 10 and 15 Myr are overplotted. In both plots the
stars confined in an area with radius of about 0\farcm6 around the
center of the association are shown with thicker symbols. The ZAMS is
drawn as a solid line. {\tt DOLPHOT} fitting errors in color and
magnitude are also shown. The PMS models predict PMS stars well above
the limit of $V\simeq$ 22 mag, but the contamination by the background
field and mostly by the close-by intermediate-age cluster BS~90 does not
allow us to distinguish the PMS stars from the evolved ones for this
magnitude range. The wide range of the isochrones which fit the loci of
the PMS stars suggests multi-epoch star
formation. However the observed broadening of the PMS stars in the CMD
might also be the result of several other factors, most prominently among them
differential reddening (see text in \S 3).}
\label{fig-multcmd} 
\end{figure*}

In the CMD a well populated faint red sequence of PMS stars is present 
(Fig.\ \ref{fig-cmdsys}). The location of the PMS stars in NGC~346 is in 
good agreement with the PMS location identified in LH~52 in the LMC 
(Gouliermis et al.\ 2006a). The faintest red part of the CMD of NGC~346, 
where these stars are located, is shown in Fig.~\ref{fig-multcmd}.  The 
whole stellar sample is plotted in this CMD, with the points corresponding 
to the stars confined within the region of the association ($\sim$ 
0\farcm6 around the center of NGC~346) being plotted with thicker symbols. 
In the left panel the tentative limits that were selected in Paper I to 
distinguish the PMS stars from other stellar types are drawn, while in the 
right panel PMS isochrone models for ages between 0.5 and 15 Myr from 
Siess et al.\ (2000) are overplotted. We assume a distance modulus of $m-M 
=18.9$ mag and, although typical value for the metallicity of the SMC is 
$Z=0.004$, we use PMS models for metallicity of $Z=0.01$, because it is 
the lowest available in the Siess et al. grid of models. A mean reddening 
of $E(B-V)=0.08$~mag (\S~\ref{sec-red}) was assumed.

\subsection{Broadening of the PMS stellar sequence \label{sc-pmsbroad}}

It has previously been suggested that the PMS stars of NGC~346 represent
a single star formation event that took place 3 - 5 Myr ago (Nota et al. 
2006).  However, a widening of the sequence of PMS stars is apparent in
the CMD of Fig.~\ref{fig-multcmd}. Comparison with theoretical
isochrones indicates that it could correspond to an age spread of
$\approx$15 Myr (Fig.~\ref{fig-multcmd}, right panel). This suggests
that the stars may not be the product of a single star formation event,
but instead star formation lasted much longer in NGC 346. In the
following, we test this hypothesis and discuss the observational
limitations and physical effects, which apart from an age spread can
produce the observed broadening of the PMS distribution. Specifically,
we analyze how photometric uncertainties, stellar variability,
unresolved binarity and variable extinction affect an observed sequence
of PMS stars.  The individual effect of each of these factors, but also
their cumulative effect on the CMD is discussed.

\subsubsection{Photometry errors}

Typical {\tt DOLPHOT} photometry errors, derived from the PSF fitting,
are shown on the right panel of Fig.~\ref{fig-multcmd} to illuminate the
dependency of the color uncertainties on the brightness of the detected
PMS stars. The errors derived from the incompleteness are significantly
lower than the photometry errors, and therefore are not considered here.
A reasonable explanation for the observed widening of the distribution
of the PMS stars could be their photometric errors. However, the
observed broadening is wider than the mean photometric errors per
magnitude range. For stars with $V$ \lapprox\ 23.5 mag the sequence of
PMS stars is found to cover a color range of at least 5$\sigma$ or more
($\sigma$ being the typical color uncertainty), while for fainter stars
with 23.5 \lapprox\ $V$ \lapprox\ 25.5 mag, this range becomes at least
3$\sigma$ wide. This implies that the photometric errors alone cannot
account for the observed spread in the whole magnitude range of the PMS
stars.

\subsubsection{Binarity}

Another factor expected to move PMS stars away from the colors and
magnitudes predicted by models is their binarity. Using the simulations
of the PMS population with $24\leq V \leq 27$~mag and $1.0 \leq V-I \leq
2.2$~mag of $\sigma$ Ori by Sherry et al. (2004), we expect at the
distance of NGC~346 ($m-M \simeq 18.9$~mag) a brightness spread of
$\Delta V\simeq$ 0.75 mag. The difference in magnitudes between the PMS
models for 1.5 and 15 Myr, for the same magnitude and color limits, is
$\Delta V\simeq$ 2.5 - 1.6 mag, at least twice as large as the spread
produced by binarity.


\subsubsection{Variability}

It is almost certain that the PMS stars of NGC~346 are T~Tauri stars, as
can be seen from a comparison of their loci in the CMD with the location
of T~Tauri stars observed, for example, in the Orion OB1 association in
the Milky Way (Brice\~{n}o et al. 2005). These stars are known to
exhibit optical variability, that could cause the PMS spread observed in
our CMD.  The variability in the $V$-band for Classical T~Tauri stars
(CTTSs) shows an amplitude up to 3 mag, while for Weak-line T~Tauri
stars (WTTSs) the variability is between 0.05 and 0.6 mag (Herbst et al.
1994). In young (3 - 5 Myr) star forming regions the ordinary fraction
for CTTSs is 30\% - 50\% (Preibisch \& Zinnecker, 1999), but this
fraction decreases with age so that by an age of a few Myr most low-mass
PMS stars are WTTSs (Sherry et al. 2004). If we assume that this region
is indeed very young, then we should consider that no more than 50\% of
the T~Tauri stars in our sample are CTTSs. These PMS stars are
characterized by irregular photometric variability (e.g. Sherry 2003)
and therefore it is quite difficult to estimate the probability of
observing them near the extreme limit of their variability. On the other
hand, the photometric variability of these objects is extremely rapid
(e.g. Sherry 2003) and consequently the probability of observing its
extreme limit is very low. Under these circumstances, a percentage of
about 50\% of CTTSs in our sample can only partly account for the
observed spread. If we consider that the region is older, as suggested
by Massey et al. (1989; $\sim$~12~Myr), then the probability that the
observed widening due to variability is even smaller, since these stars
would be mostly WTTSs.

\subsubsection{Reddening}

Differential reddening is probably the most important factor that might
produce the observed broadening. In \S~2, we derive the individual
interstellar reddening for each of the OB stars within the association. 
This leads to an estimation of their color excess $E(B-V)$ and
consequently their individual extinction $A_{\rm V}$ from their
un-reddened color indices $(B-V)_{\rm 0}$, and reddening-free $Q$
parameters (\S~\ref{sec-difred}; Table \ref{tab1}).  We find that
extinction varies from $A_{\rm V} \simeq$ 0.06 to 0.39 mag (with a peak
at $A_{\rm V} \sim$ 0.13 mag). Such an extinction-spread corresponds to
a spread in color excess between $E(V-I) \simeq$ 0.03 and 0.20 mag,
which is much smaller than the observed widening in color of the
sequence of PMS stars. On the other hand, if we consider the reddening
distribution based on the upper MS stars of the entire observed region
(Fig.~\ref{fig-genred}; $\langle E(B-V)\rangle \sim 0.08$~mag), then
reddening becomes a very important factor for the observed PMS
broadening. A further support to the crucial role of reddening is given
by the comparison of the spatial variations in the locus of the PMS
stars with the reddening map of the observed region derived in
\S~\ref{sec-red} (shown in Fig.\ref{fig-genred} left) and the spatial
variations of the reddening of the central 31 OB stars
(\S~\ref{sec-difred}). This comparison shows that indeed the PMS stars
located in the more reddened areas seem to occupy a preferentially
redder locus in the CMD. The importance of reddening is demonstrated
below with the use of simple simulations of the observed loci of PMS
stars in the CMD.


\subsection{Simulation of the Observed PMS Broadening
\label{sec-simbroad}}

We discussed above several factors that may explain the observed PMS
spread in NGC~346, and we find that none of them can actually produce
this spread in the observed degree {\em if it acts alone}. However, one
should consider that these factors are not isolated phenomena, but, on
the contrary, they are expected to act on the observed stellar
parameters in a cumulative manner. Taking this into account, we simulate
the observed widening of the PMS assuming that the ``suspected'' factors
act in an additive way on a sequence of single-age PMS stars. We
constructed a toy model of the observed PMS stars, assuming that all
stars are the product of a single star formation event. Then, we applied
each of the aforementioned biases. We conclude from this test that a
single-aged PMS population can indeed be misinterpreted as the product
of star-formation events that took place during different periods. The
apparent age spread can be the product of the cumulative action of
several factors. 

An example of the results of our toy model is given in
Fig.~\ref{fig-sim}, where we assume that the observed PMS population of
NGC 346 is 4 Myr old.  First, we assume a concentration of PMS stars, as
they are selected in Paper I according to their location in the CMD
(Fig.~\ref{fig-sim}{\em a}), which populates a single 4~Myr PMS
isochrone (Fig.~\ref{fig-sim}{\em b}). Then we apply the effects of the
suspected factors one on top of the other. Here we describe the
procedure:

\begin{itemize}

\item[(i)] {\em Binarity}: As far as the existence of unresolved
binaries in our sample concerns, there are no studies on the expected
fraction of binary systems in associations of the Magellanic Clouds.
Therefore, we based our simulations on the information we have on
galactic OB associations. These systems show a binary fraction that does
not differ significantly from the field (Mathieu 1994), which is found
for G dwarfs to be around $\sim$ 58 \% (Duquennoy \& Mayor 1991). The
well-studied Orion Nebular Cluster (ONC) is also found to be consistent
with the field binary fraction (e.g. Prosser et al. 1994; Petr et al.
1998; K{\"o}hler et al. 2006). In many clusters, a clear binary sequence
is observed to lie 0.75 magnitudes above the single star main sequence
(Tout 1991), and this is almost equal to the brightness shift that a
single star will suffer toward brighter magnitudes, if it is an
unresolved binary system with equal mass components (de Bruijne et al.
2001).

Simulations of unresolved low-mass PMS stars in the Orion OB1b association 
centered on $\sigma$ Ori by Sherry (2003) showed that for a 3 Myr old 
binary system with a 0.5 M{\solar} primary the position of the system on 
the $V-I$, $V$ CMD can be shifted up to 0.13 mag on the $V-I$ axis. The 
actual shift depends on the mass of the lower-mass companion, which ranges 
from 0.1 to 0.5 M{\solar}. This system also brightens by slightly less 
than 0.75 magnitudes. It should be noted that recent observational studies 
on the mass ratios of T~Tauri binaries are consistent with a uniform 
distribution of mass ratios between 0.2 and 1.0 (Woitas et al. 2001). In 
order to simulate the effect of unresolved binaries to the modeled 
sequence of 4 Myr old PMS stars we randomly applied a brightening of no 
more than $V \simeq$ 1 mag and a reddening \lapprox\ 0.13 mag, assuming 
different binary fractions. It is worth noting that Elson et al. (1998) 
measured a binary fraction in the young rich LMC cluster NGC~1818, and 
found $35 \pm 5$\% in the core. This is roughly comparable to the Galactic 
values which are considered here. In Fig.~\ref{fig-sim}{\em c} (top panel) 
the result for a binary fraction of 50\% is shown. This result clearly 
suggests that binarity can play an important role to the broadening of a 
single-age PMS population in the $V-I$, $V$ CMD.

\item[(ii)] {\em Variability:} In order to simulate the effect of
variability on our modeled sequence of PMS stars, we assigned to each
star a random amplitude drawn from a Gaussian probability distribution,
following Sherry (2003), who assumed normally distributed $V$-band
variations in WTTSs with a mean of $M_{V} = 0.15$ mag and a standard
deviation of $\sigma =$ 0.1 mag, based on data for 42 WTTSs with known
$V$-band variations. The $V-I$ colors of these stars should appear
redder when they are fainter, due to rotational modulation. According to
Herbst et al. (1994) the ratio of color variations, $\Delta I/\Delta V$,
ranges from 0.39 to 0.88 mag. We assumed several values for this ratio
within these limits. In Fig.~\ref{fig-sim}{\em d} (top panel) we show
the results assuming a mean value of $\Delta I = 0.6 \Delta V$ and
consequently $\Delta(V-I)=0.4 \Delta V$. It should be noted that for
these results we assumed that only WTTSs are present at the assumed age
of NGC~346. This is certainly the reason for a rather small effect on
the variability in our simulations. In the case that some fraction of
CTTSs had also been included, the spread implied by the model could be
as large as 5 times more, due to the higher amplitude these stars
exhibit.

\begin{figure*}[]
\epsscale{.75}
\plotone{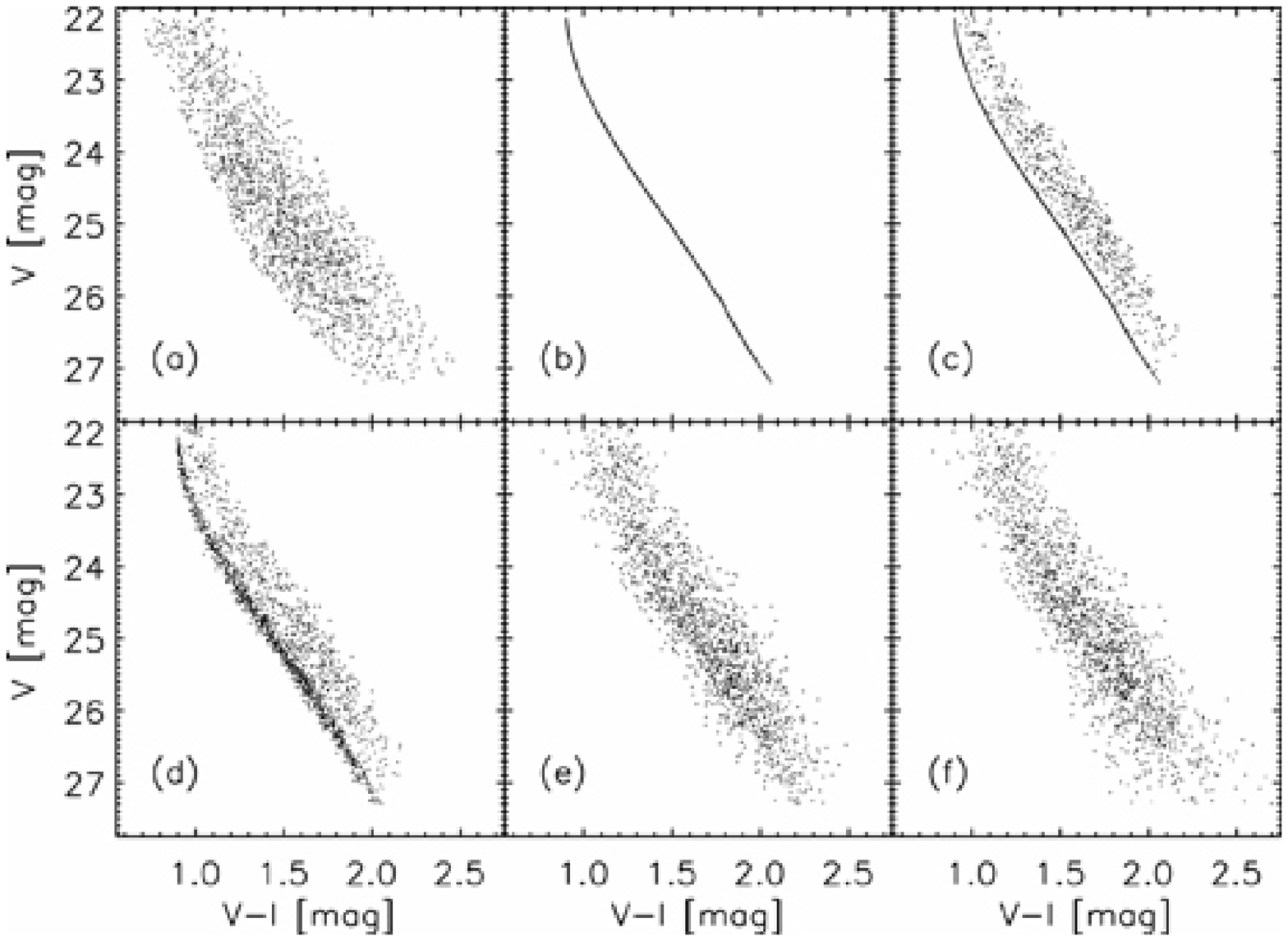}
\plotone{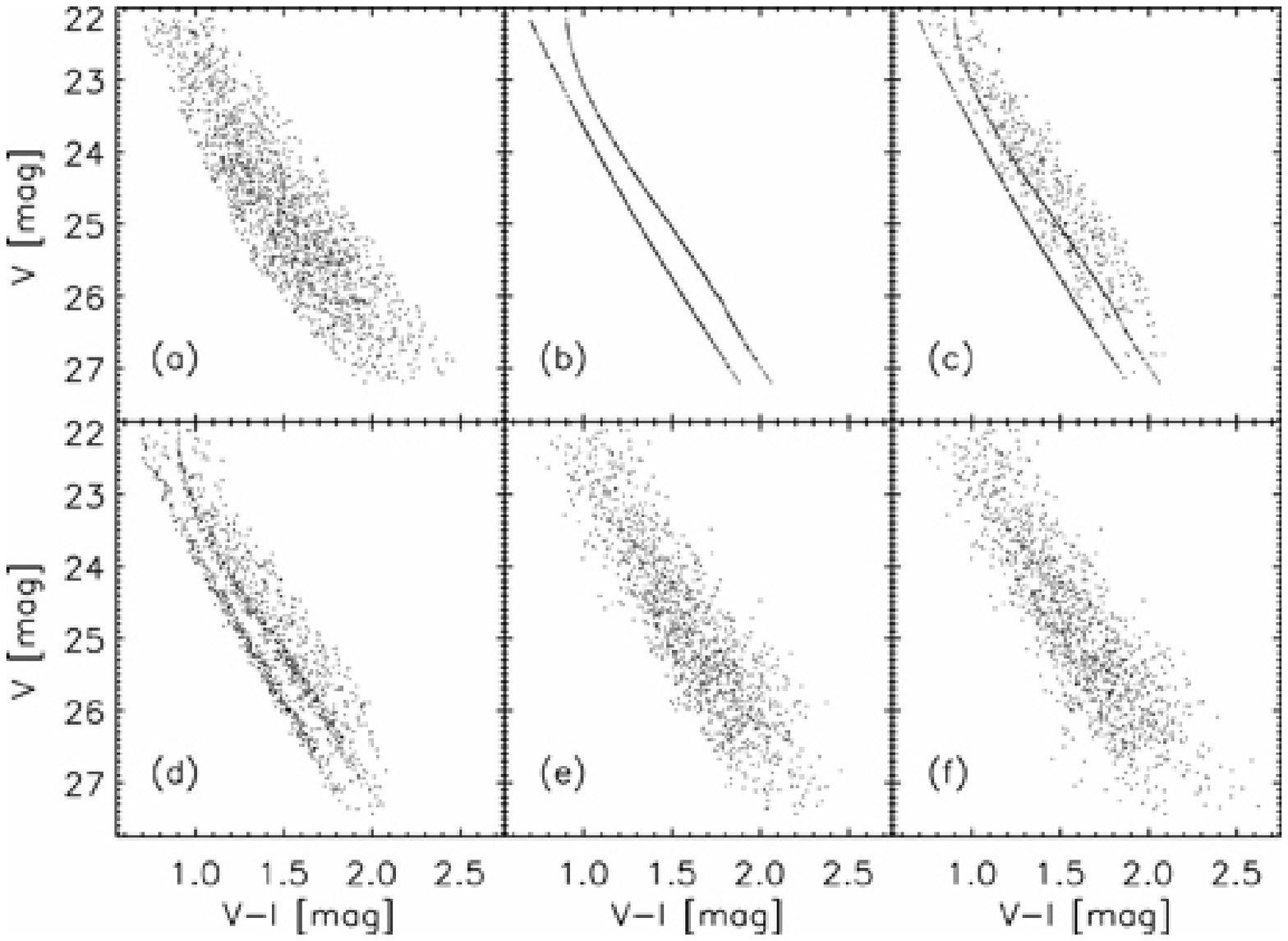}
\caption{{\em Top}: An example of the results from our toy model used to 
simulate the cumulative effect of several factors on a single-age sequence 
of PMS stars. (a) The observed PMS stars in the region of NGC~346 are 
selected according their positions in the $V-I$, $V$ CMD. (b) Their 
positions are modeled based on the 4~Myr PMS isochrone from Siess et al. 
(2000). Factors, such as (c) binarity, (d) variability, (e)  reddening and 
(f) photometric uncertainties are applied in order to explore the 
broadening they cause to a single-age PMS population, making it appearing 
as the result of multi-epoch star formation (see \S~\ref{sec-simbroad}). 
{\em Bottom}: Results from a toy model similar to the one on the top, but 
assuming that the PMS stars are formed during two star formation events, 
one 10~Myr and the other 4~Myr ago. In this example 50\% of the population 
is assumed to be formed during the first and 50\% during the second 
event. These results imply that the possibility of more than one star 
formation events in NGC~346/N~66 cannot be ruled out.}
\label{fig-sim} 
\end{figure*}

\item[(iii)] {\em Reddening}: We explore the displacement of the PMS stars 
from their theoretical loci on the CMD due to reddening. We assume that 
these stars are subject to a) the reddening of the whole area as estimated 
from the upper main sequence stars (\S~\ref{sec-red}), by randomly 
applying the observed reddening distribution shown in Fig. 
\ref{fig-genred} ({\em right}), and b) the much lower reddening of the 
area of NGC~346 alone found from the OB stars (\S~\ref{sec-difred}). For 
Fig.  \ref{fig-sim}{\em e} (top panel), the assumed reddening distribution 
was based on stars in the whole region with $V$ \lsim\ 21 mag, and is 
chosen to show the maximum effect. The corresponding distribution for even 
brighter stars is thinner with smaller mean values and standard 
deviations. This would also be the result for the (more accurately 
measured) reddening found for OB stars. As shown in Fig. \ref{fig-sim}{\em 
e}, high extinction is indeed the most important factor, in addition to 
binarity, that can cause the observed broadening of single-age PMS stars.

\item[(iv)] {\em Photometry}: Photometric accuracy is a function of 
magnitude, being higher for the brighter stars. In order to explore the 
effect of photometric uncertainties to a ``perfect'' sequence of PMS 
stars, we randomly applied the observed mean photometric errors per 
magnitude (shown in Fig.~\ref{fig-multcmd} {\em right}) to the modeled PMS 
population. Naturally, the derived broadening is wider for the fainter 
stars. This is shown in Fig.~\ref{fig-sim}{\em f} (top panel).

\end{itemize}

\begin{figure*}[]
\epsscale{.75}
\plotone{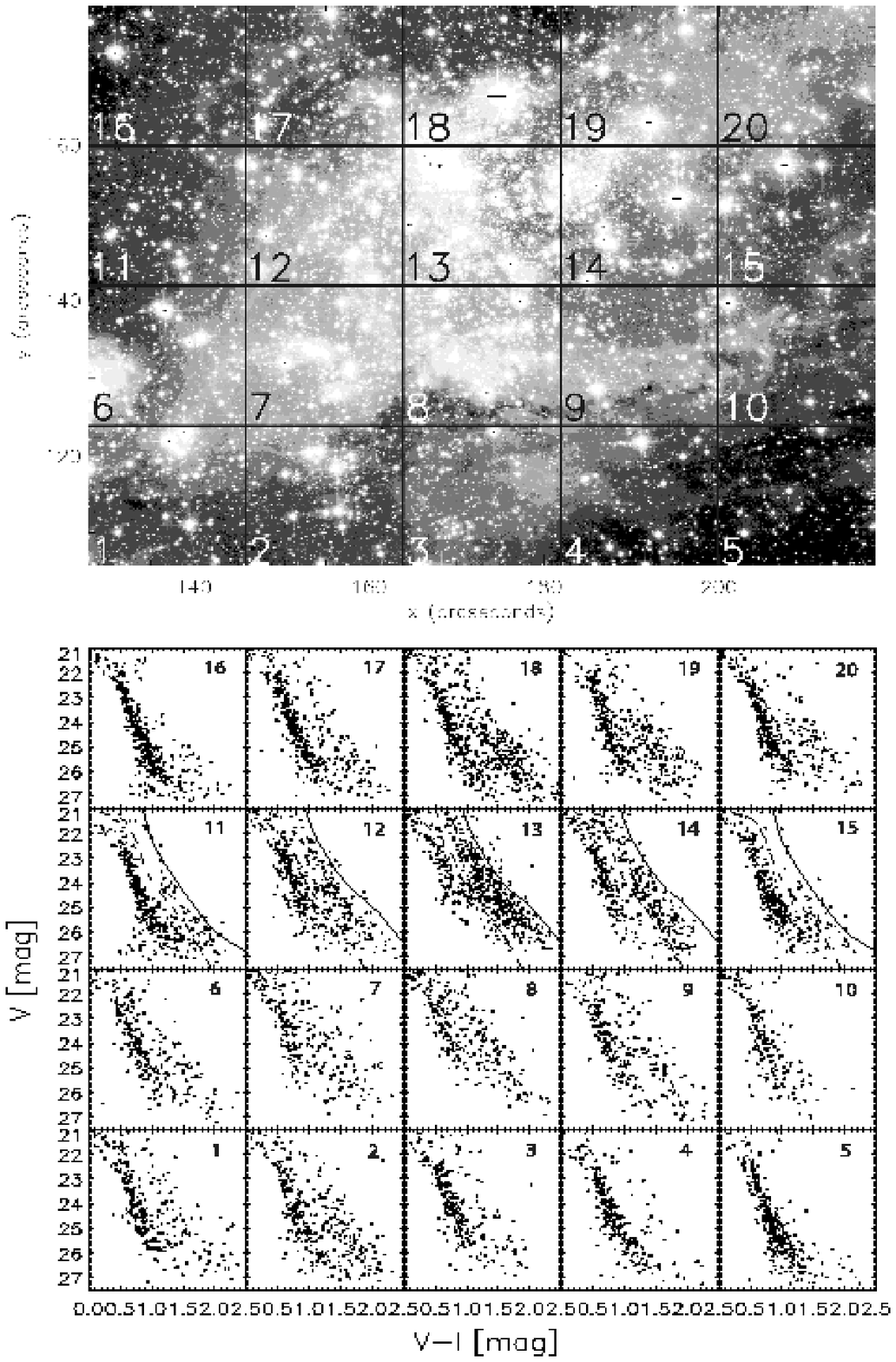}
\caption{{\em Top}: A 1\farcm5 $\times$ 1\farcm2 region centered on the
association NGC~346, with the borders of 20 selected sub-regions
18\arcsec$\times$18\arcsec\ wide each overplotted.  {\em Bottom}: The
CMDs of the sub-regions, with PMS isochrones overplotted for sub-regions
11 to 15.  PMS stars are present in almost the entire region,
and they seem to be aligned across the nebula. This plot shows that the
CMD varies in number of PMS stars and width of their broadening from one
region to the other.} 
\label{fig-cntr346} 
\end{figure*}

From the results of our toy model for the influence of several factors
on a ``perfect'' single-age sequence of PMS stars, it can be seen that
indeed a single-age PMS population may appear broad enough in the $V-I$,
$V$ CMD so that it could be misinterpreted as a group of stars, produced
by multi-epoch star formation. We specifically note the importance of
reddening and binarity of such stars for this broadening.  Variability
may also play an important role depending on the nature of these stars. 
Although each of these factors acting alone cannot cause the observed
widening of PMS loci in the CMD, it is their cumulative action that
leads to the broadening of the sequence of PMS stars in NGC 346.
However, it should be noted that if indeed reddening is the most
important cause for this widening, then the PMS population of NGC~346
should be expected to be older than 3 - 5 Myr, since reddening
``pushes'' PMS stars to redder colors making them appear {\em younger}.
Specifically, PMS models for ages on the order of 10 Myr proved to fit
the observed sequence of PMS stars and their broadening better. This
result is in line with Massey et al. (1989), who claim that although the
location of the bright stars of NGC~346 in the CMD implies an age of
less than 5~Myr for the association, there is a spatial distinct
subgroup of evolved $\sim$~15~M{\solar} stars {\em in the association}
with ages $\simeq$~12~Myr.

Furthermore, another very interesting result of our simulations is that 
the possibility of multi-epoch star formation cannot be excluded. 
Specifically, our toy model showed that if we consider that the 
interstellar reddening may be lower than what was previously assumed 
(using e.g. $E(B-V)\simeq0.04$~mag found from the OB stars in the region 
of the association), then the observed spread of the PMS stars can only be 
reproduced by (at least) two separate PMS isochrones. This is demonstrated 
in Fig.~\ref{fig-sim} (bottom panel), where we show the results of the toy 
model if two star formation events around 4 and 10 Myr ago are assumed. In 
this figure it is shown that the final broadening of the PMS loci in the 
CMD does not differ significantly (although it is a bit wider) from the 
one resulted from a single star formation event. Typical ages derived from 
the simulations of a double star formation event (half of the stars formed 
in each epoch) are 3 - 4 and 10 - 12 Myr in agreement with both 
hypotheses.  Conclusively, we {\em cannot} completely exclude the 
possibility of a true age spread of the sequence of PMS stars in NGC~346 
due to separate star formation events.

\subsection{CMD Variations within the Region of NGC~346\label{sec-cmdvar}} 

In order to study variations in the CMD we selected a region of size
1\farcm5 $\times$ 1\farcm2, centered on the association NGC~346. This
region was divided into 20 18\arcsec$\times$18\arcsec\ sub-regions. 
Indeed, we see that the CMDs differ from one region to the next in the
nature of the stars, their numbers and the apparent PMS broadening. The
individual CMDs of all sub-regions are shown in Fig.~\ref{fig-cntr346}.
In order to avoid confusion, PMS isochrones are overplotted only for
sub-regions 11 to 15, so that a rough estimation of the age differences
can be extracted. Specifically, isochrones for 1.5 and 10 Myr are
displayed in the CMD of sub-region 11, the 0.75 and 15 Myr isochrones
are shown in the CMDs of sub-regions 12 and 13, whereas for the CMD of
sub-region 14 the older age limit seems to be better fit by a 10 Myr
isochrone, but for a very low number of faint PMS stars.

From Fig.~\ref{fig-cntr346}, it can be seen that PMS stars are present
in almost the entire selected region, except for the southwestern
(sub-regions 5 and 10) and northeastern (sub-region 16) areas, which are
located away from the association (and the nebula). From the CMDs of the
remaining sub-regions we can conclude that PMS stars seem to follow the
southeast (bottom left corner) to northwest (top right corner) direction
of the \emph{nebulosity}, with a higher concentration in the main body
of the association (sub-region 13) and around it (e.g. sub-regions 12,
14 and 18)\footnote{The concentration of PMS stars in sub-regions 3, 4,
5, 10, 11, 15, 16 and 17 is lower than the one along the
southeast-northwest axis.}. Reddening in the selected region is expected
to be higher in the part centered on the main body of NGC~346 due to the
nebula, and indeed the corresponding CMDs exhibit the wider broadening
in their PMS populations. This suggests that reddening {\em is} indeed
(at least partly) responsible for the observed PMS broadening. This is
more clearly seen in the CMD of sub-region 13, the best representative
of the population of the association (similar to the CMD used in our
toy-model).

From the discussion so far on the PMS stars of association NGC~346 alone
we {\em cannot} safely conclude that these stars represent a coherent
population, formed in a single star formation event. Nota et al. (2006)
and Sabbi et al. (2007) argue that such an event happened 3 - 5 Myr ago,
while Massey et al. (1989) suggest a timescale three times longer,
closer to our results presented in \S 3.3. However, taking into account
the low reddening found from the OB stars in the association, we tend to
conclude that a true age spread may also contribute to the observed PMS
broadening.

NGC~346 does not seem to represent the only product of recent star
formation in the general area.  As shown from the isodensity contour
maps of Fig. \ref{fig-cms}, density peaks of PMS stars can be seen away
from the association, forming small compact concentrations. In the
following sections we study these concentrations, which we refer to as
``PMS clusters''. We also present our results from H\alp\ observations
and previous results from observations with the {\em Spitzer} Space
Telescope in the mid- and far-infrared, which reveal loci of on-going
star formation in the general region of NGC~346/N~66.

\section{PMS Clusters in the region of NGC~346/N~66 \label{sec-pmsclus}}

\subsection{The most significant PMS clusters in the Region}

\begin{figure*}[t!]
\centerline{
\vbox{
\psfig{figure=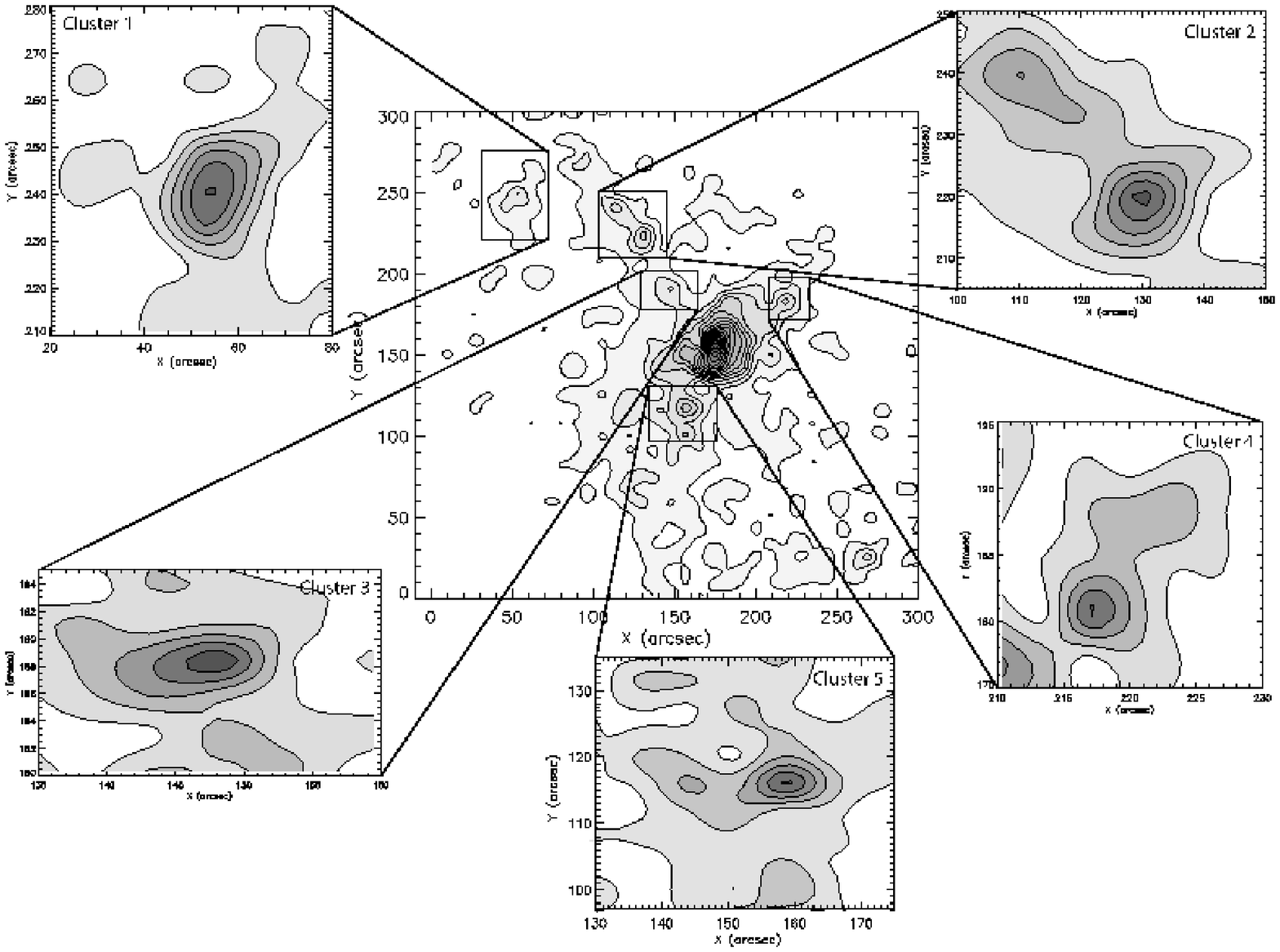,width=12cm,angle=270}
\psfig{figure=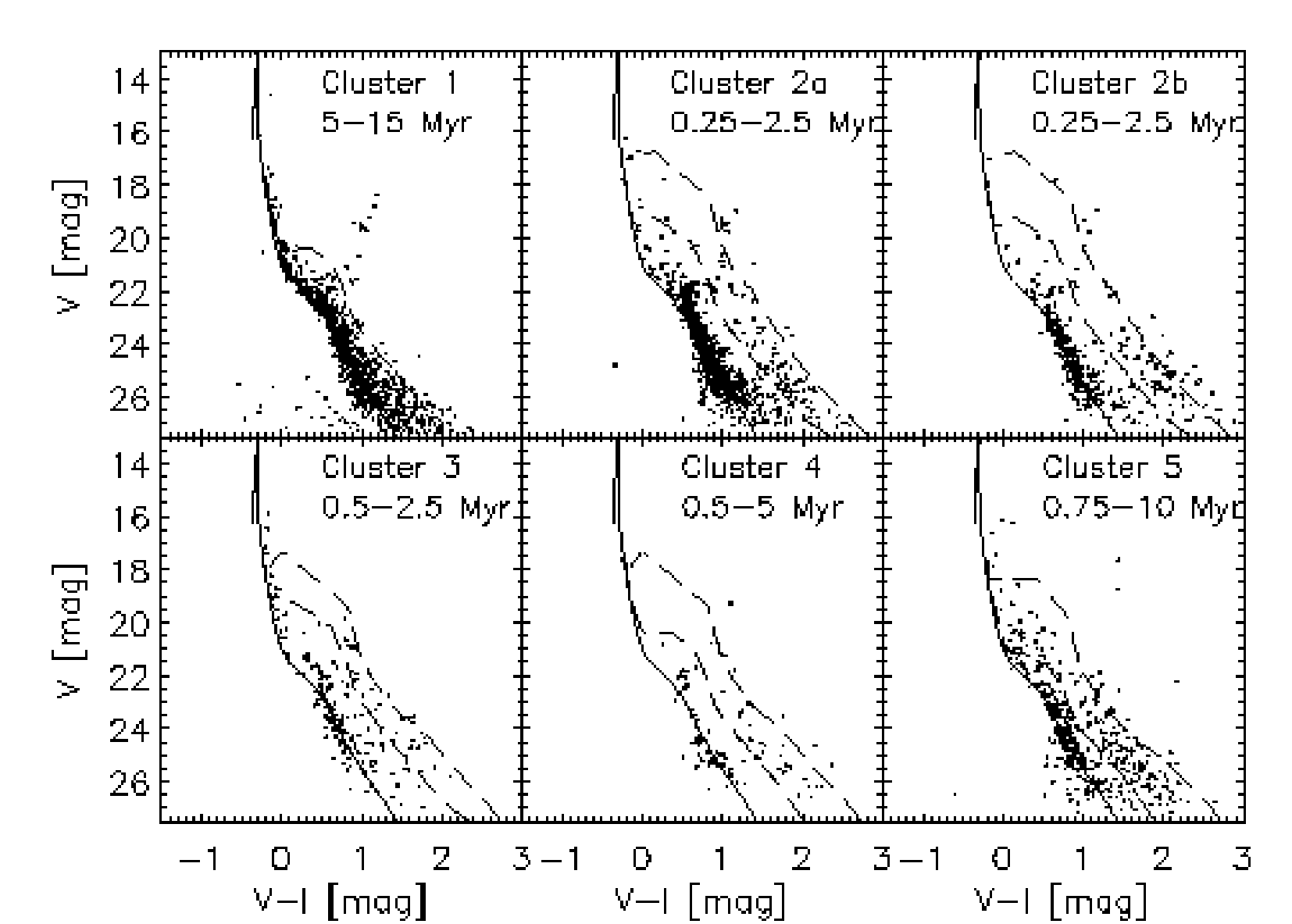,width=13cm}
}
}
\caption{Top: Isodensity contour map from star counts of the PMS stars of the 
area of NGC~346/N~66, as they have been selected in Paper I. Five PMS 
clusters in the area around the association NGC~346 are selected based on 
their high density of PMS stars and the corresponding local contour maps 
are shown enlarged. Bottom:
$V-I$, $V$ CMDs of the stars included within the area of each
of the selected PMS clusters.
An isochrone for an age of $\approx$4\,Myr by Girardi et
al.\ (2002) is overplotted for the main sequence populations. For the PMS
stars, isochrone models by Siess et al. (2000) are overplotted to
indicate the time range within which these stars have presumably formed.
Cluster 1 seems to be rather a more evolved open cluster, while cluster
5 meets the characteristics of the association itself.  Clusters 2a, 2b
and 3 probably represent the youngest products of clustered star
formation away from the association NGC~346.
\label{fig-pmsclus}}
\end{figure*}

The isodensity contour map of the region of NGC~346/N66 constructed from
star counts of the PMS stars is shown in Fig.~\ref{fig-cms} (right
panel).  In this map several distinct concentrations of PMS stars can be
identified. We selected the statistically most significant
concentrations with a density \gsim\ 3$\sigma$ above the background
(where $\sigma$ is the standard deviation of the background density).
There are five such concentrations, which fit the description of ``PMS
clusters''. The contour map of the whole region, as well as the ones of
the local regions around each of these clusters, numbered from north to
south, are shown in Fig.~\ref{fig-pmsclus} (top).

\begin{figure}[]
\epsscale{1.15}
\plotone{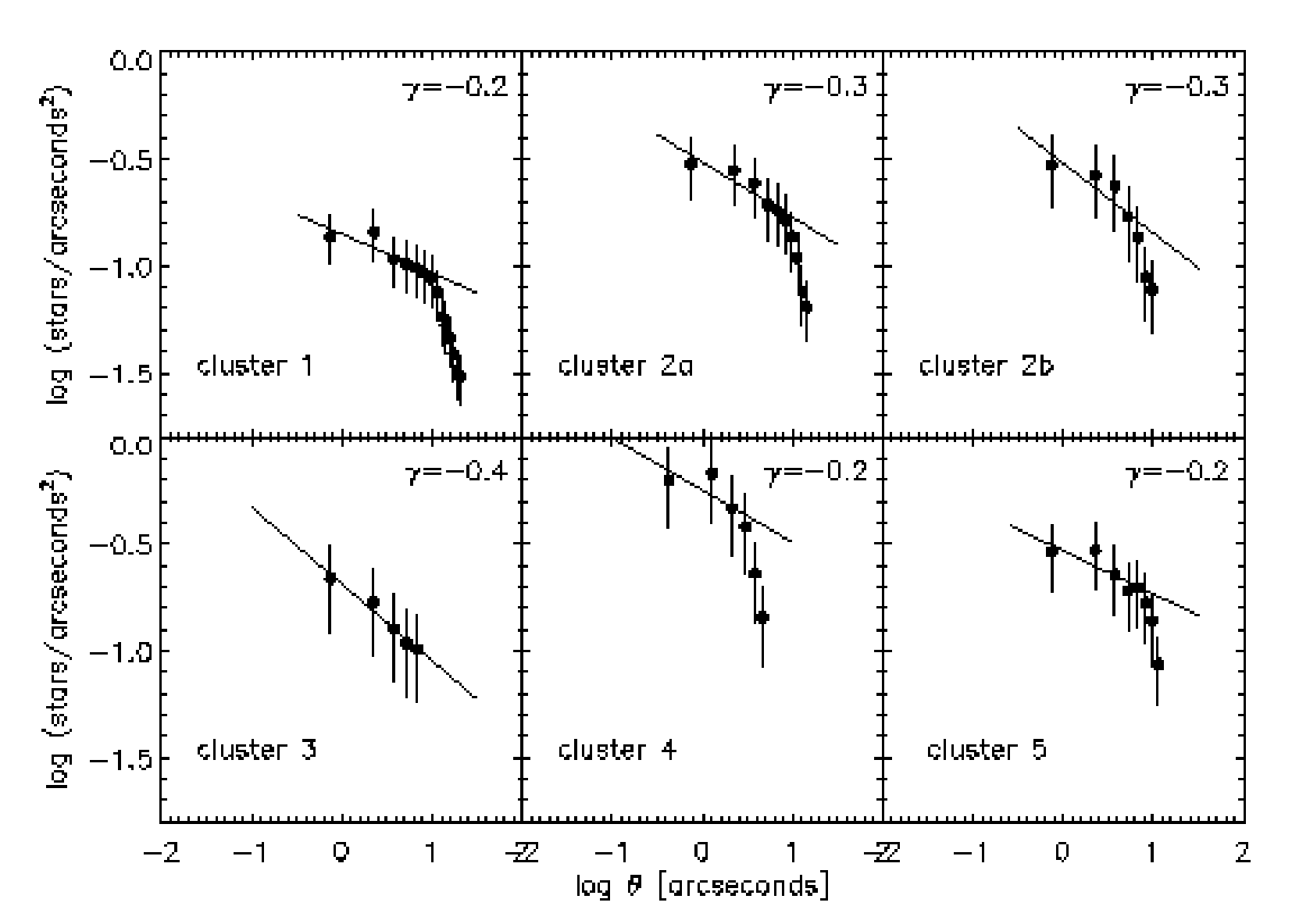}
\caption{Average surface density of companions as a function of angular
separation from each star for the PMS clusters identified in the region
of NGC~346 (Fig.~\ref{fig-pmsclus}). From our observations we cannot
distinguish separations in the binary regime. Therefore, the power-law
index is given only for larger separations, typically in the range
0.8$''$ to 8$''$, corresponding to physical scales of 0.24\,pc to 2.4\,pc at
the distance of the SMC. The fitting functions, presented
by solid lines, and the corresponding indices, $\gamma$, are given for
each cluster. Their values indicate that all PMS clusters should be
considered as results of a hierarchical clustering process (see section 4.2).} 
\label{fig-clsd} 
\end{figure}

The size of each PMS cluster was defined by a closer inspection of their
contour map, as shown in Fig.~\ref{fig-pmsclus} (top). The center was
chosen to be the density peak. A circular area was selected to represent
the extent of each cluster with a radius determined by the isopleth,
corresponding to the mean density of the surrounding region (first
isopleth in the individual maps of the clusters). This isopleth defines
the size of each cluster. It should be noted that cluster 2 consists of
two distinct regions, which we will treat in the following study
separately. Clusters 1 and 3 are clearly detected in the contour maps of
both PMS and Upper Main Sequence stars (see Fig.~\ref{fig-cms}).
Therefore, their size was determined by the contour maps constructed
with both PMS and UMS stars. In order to define an age range within
which each PMS cluster was presumably formed we construct the CMD of the
stars included in the area covered by each cluster.

\subsubsection{Color-Magnitude Diagrams of the PMS Clusters}

Although it is quite difficult to disentangle the actual age of the
discovered PMS clusters, because of the artificial broadening of the PMS
loci in the CMD discussed in \S~\ref{sec-pms}, it is worthwhile to
specify the limits within which their ages may range, in order to
identify any age-differences between them. Therefore, the stellar
populations of the PMS clusters are analyzed in terms of their CMDs
shown in Fig \ref{fig-pmsclus} (bottom). Isochrone models for both MS
(Girardi et al. 2002) and PMS (Siess et al. 2000) populations are
overplotted. These models suggest that cluster 1 is in a rather evolved
state with the best fitting PMS model indicating an age range of 5 to 15
Myr. The stellar content of this cluster, which also includes a few
bright MS stars, suggests that it should be treated as a small open
cluster rather than a recently formed PMS cluster. 

The models indicate that all the remaining PMS clusters seem to be much 
younger than cluster 1.  Specifically, clusters 2 to 4 do not exceed an 
age of 5 Myr, with clusters 2a, 2b and 3 being the youngest in the sample 
with ages for their PMS stars younger than 2.5 Myr. All isochrone models 
have been plotted for a mean reddening of $E(B-V)\simeq$~0.08~mag. 
Differential reddening is found to play an important role in the 
broadening of the PMS stars in the CMD. However, a constant mean reddening 
should be considered for the isochrone models, and we found that the 
selection of this reddening according to the measurements of 
\S~\ref{sec-red} and \S~\ref{sec-difred} does not seem to be very 
important for the derived age-span of the clusters. This is demonstrated 
in Fig \ref{fig-pmsclus} for the CMD of cluster 2a, where we plot the PMS 
isochrones assuming three cases: No reddening (dash-dotted lines) and 
reddening values of $E(B-V)\simeq$~0.08~mag (found from the upper main 
sequence for the whole region; dashed lines) and $E(B-V)\simeq$~0.04~mag 
(found from the OB stars in the association; long-dashed lines).

The spread of the PMS stars in the CMD of cluster 5 covers a wider range 
of ages, comparable to the one we observe in the main body of the 
association NGC~346 (discussed in \S 3). This is not a surprising result 
considering that cluster 5 is the closest PMS cluster to the association, 
and probably it shares the same star formation history with it. In 
general, from the individual CMDs of the PMS clusters (except for cluster 
1, which is a rather evolved open cluster with a PMS population) we see 
that the northern part of the observed field (north of the association) 
covers younger clusters of PMS stars than in the southern part and the 
association itself. It will be discussed below whether this is an 
indication that these clusters do not share the same star formation 
history with NGC~346, but rather that they are the products of more recent 
star formation.

\subsection{Cluster Properties of PMS Stars}

The degree of clustering of PMS stars can be evaluated through their
surface density as a function of angular separation from each star.
Gomez et al. (1993) applied this technique to the study of the spatial
distribution of PMS stars in the Taurus-Auriga molecular cloud, and they
found that a power law form (with index $\gamma \simeq -1.2$) reproduces
the overall shape of the actual PMS distribution in Taurus, indicating
the presence of real clustering. Larson (1995) extended the analysis of
Gomez et al. to smaller separations and he measured a power-law relation
$\Sigma \propto \theta^{\gamma}$ for young star clusters, where $\Sigma$
is the surface density of PMS stars, and $\theta$ the angular separation
from each star. He found that the Taurus young stars exhibit
self-similar or fractal clustering on the largest scales, but there is a
clear break from self-similarity at a scale of about 0.04 pc which
divides the regime of binary and multiple systems on smaller scales from
that of true clustering on larger scales.

A survey of young clusters in the Orion, Ophiuchus, Chamaeleon, Vela,
and Lupus star-forming regions by Nakajima et al. (1998) showed an index
for $\Sigma(\theta)$ of $\gamma \simeq -2$ at stellar separations in the
binary regime, smaller than 0.1 pc, whereas at large separations (0.1 to
1.0 pc.), the power index seems to vary from region to region with $-0.8
< \gamma < -0.1$.  The break in the power law was found to occur on
physical scales of 0.01 to 0.1 pc. Pudritz (2002) argues that this
spatial range probably does not reflect the initial conditions for
isolated star formation in clusters, because a star moving at $\simeq
1.0~{\rm km} {\rm s}^{-1}$ in such a cluster will move a distance of 1.0
pc in a million years, thereby erasing the initial conditions. According
to Larson, the power law break provides clear evidence for an intrinsic
scale in the star formation process, which is found to be essentially
equal to the Jeans length in typical molecular cloud cores. He also
notes that there is evidence that nearly all stars are formed in binary
or multiple systems, and that some of these systems are subsequently
disrupted by interactions in the denser star-forming environments to
produce the observed mixture of single, binary and multiple stars
(Larson 1995). Goodwin et al. (2007) review the observed properties of
multiple systems and find that the multiplicity of stars changes,
probably due to the dynamical decay of small-N systems and/or the
destruction of multiple systems within dense clusters. They note that
models of the fragmentation of rotating and turbulent molecular cores
almost always produce multiple systems, but the properties of those
systems do not match the observations.

OB associations are known to contain subgroups of smaller clusters
(Blaauw 1983), which often contain smaller groups of stars themselves,
made of multiple-star systems. Efremov \& Elmegreen (1998) suggest that
star formation is generally clumped into a hierarchy of structures, from
small multiple systems to giant star complexes and beyond and that OB
associations are just one level in this hierarchy. Even the ISM has been
observed to be arranged in a hierarchical structure (Scalo 1985;
V{\'a}zquez-Semadeni 2004). Turbulent motions have been proposed as a
cause for the hierarchical structure (Elmegreen 1998).  The velocity
structure in turbulent regions is suggested to be scale-free, meaning
that {\em the same physical processes and the same velocity-separation
relations occur over a wide range of absolute scales} (Elmegreen 2006). 
According to this hypothesis, large clumps originate from the large
velocities which compress the gas on large scales and small clumps
result from the compression of the gas characterized by small velocities
on small scales. Considering that stars are formed by the collapse of
such hierarchical structured clumps, they are themselves hierarchically
structured. The break in the power-law relation $\Sigma \propto
\theta^{\gamma}$ defines the two different regimes of star formation:
The binary regime at smaller scales, and the regime of hierarchical
clustering at larger ones.
 
In order to determine the clustering properties of the PMS clusters
found in the region of NGC~346, we apply the same technique and
construct the average surface density of companions $\Sigma(\theta)$ for
every PMS member of each of the PMS clusters as a function of the
angular projected separation $\theta$ (see Fig.~\ref{fig-clsd}). We
apply a linear fit for the estimation of the power law index, $\gamma$.
The function $\Sigma(\theta)$ drops sharply near the largest separations
owing to the size of the region selected for each cluster. Consequently,
our analysis is based only on the smaller observed separations.  In
addition our observations do not allow us to separate stars in the
binary regime and, therefore, our study does not include them. The
errors shown in Fig.~\ref{fig-clsd} reflect the low number statistics in
our samples of PMS stars per cluster.

The derived values of the power-law index for all clusters are found to
be in the narrow range $-0.4 < \gamma < -0.2$, comparable to the results
of Nakajima et al.  (1998) for separations in the regime of hierarchical
clustering. The power-law index relates to the nearest-neighbor
distance, in the sense that if the latter can be fitted by the Poisson
distribution, then the power index is close to 0. The distribution of
the nearest-neighbor distance is much broader than the Poisson one, when
the surface density varies appreciably within the region. Then
$\Sigma(\theta)$ has a steep $\theta$-dependence, and the power-law
index is negative (Nakajima et al. 1998), comparable to the values we
find for our clusters.  Consequently, these values can be interpreted as
the result of the variation of the surface density within the region,
and clearly suggest that the PMS clusters identified in the region of
NGC~346/N~66 are the products of hierarchical clustering.


\begin{figure*}[t!]
\epsscale{1.15}
\plottwo{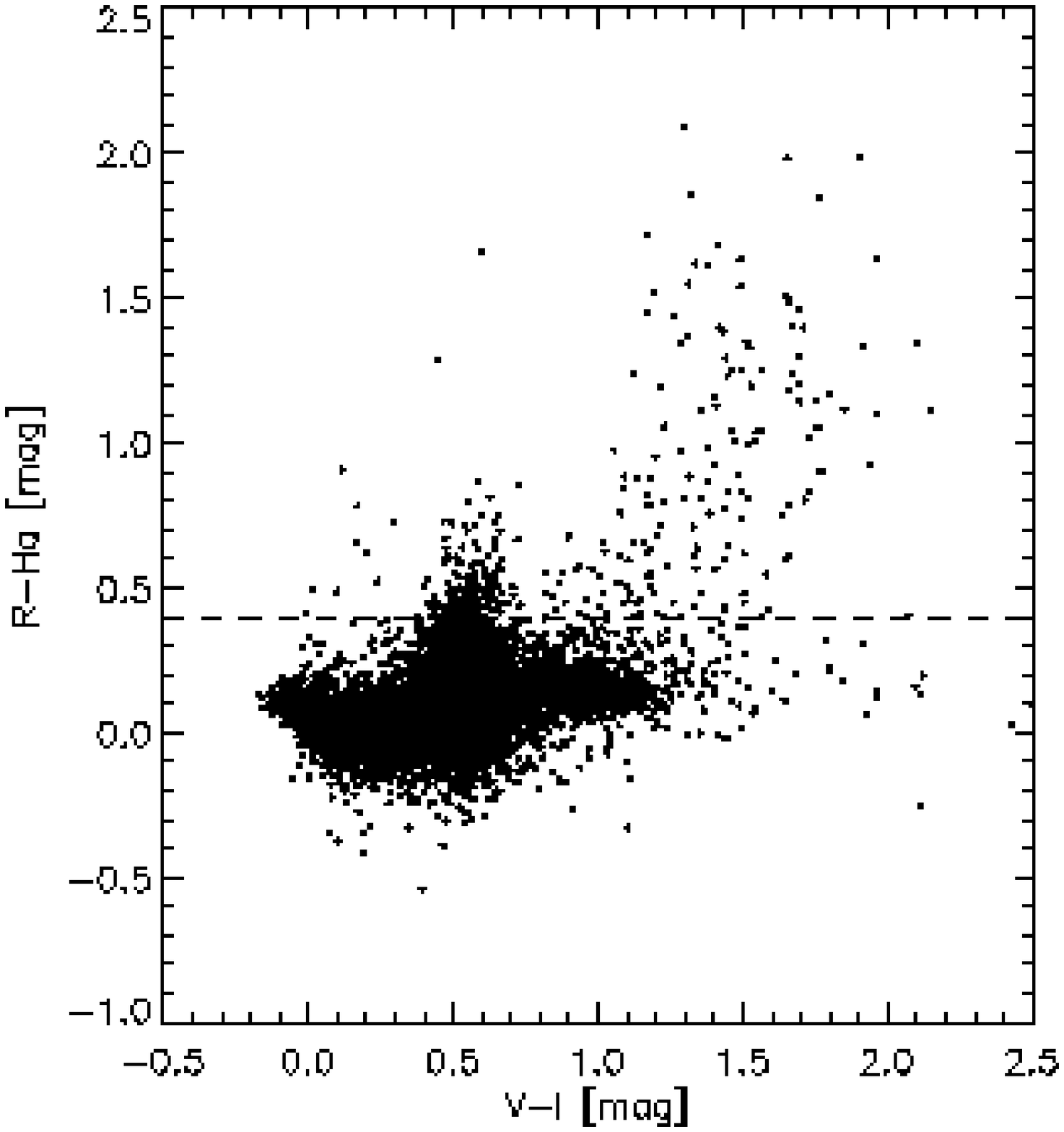}{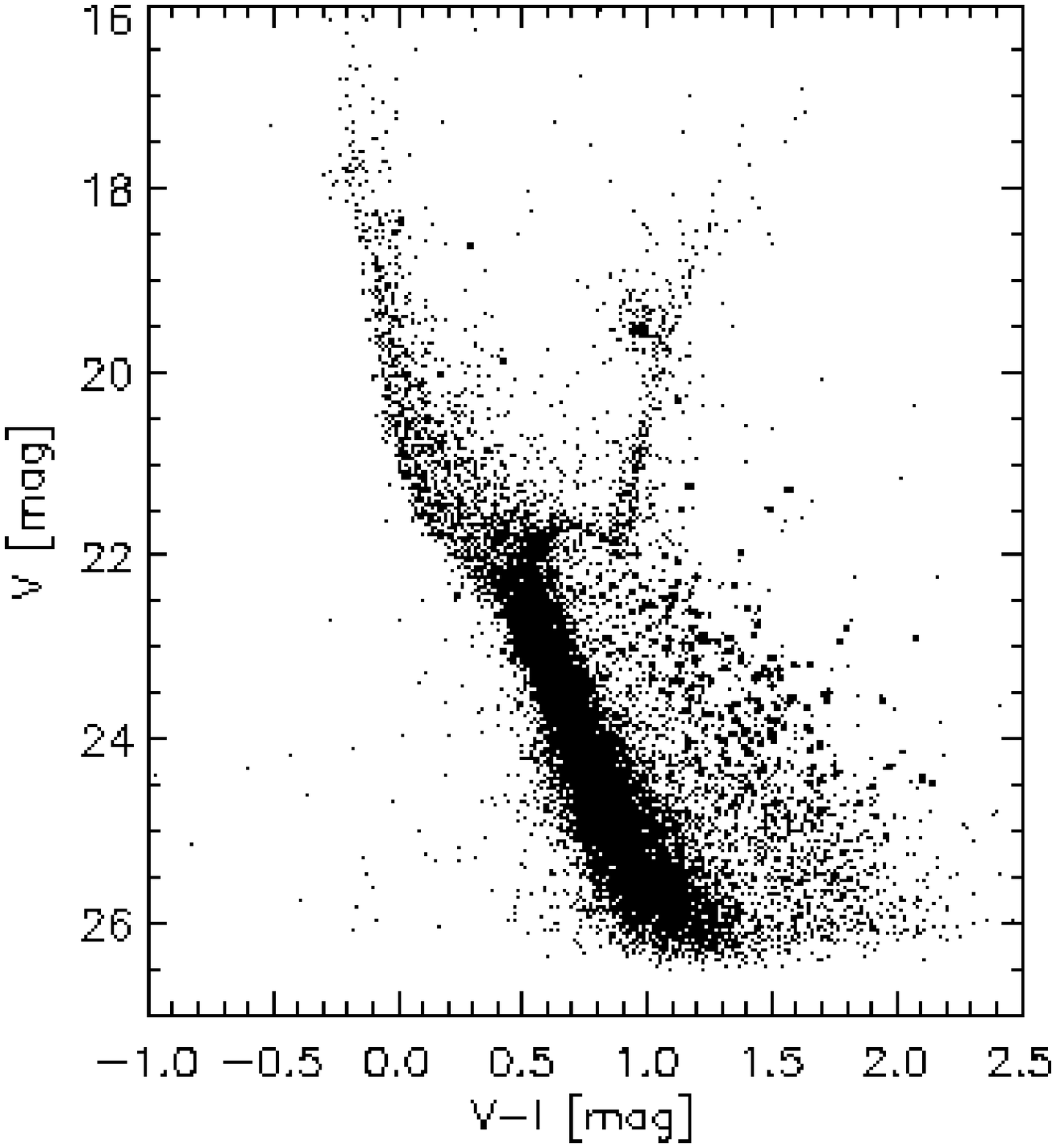}
\caption{{\em Left}: In a $R-{\rm H}\alpha$, $V-I$ color-color diagram
stars without  H\alp\ emission form an almost horizontal band around $R-{\rm
H}\alpha \simeq 0.0$ mag, with most of the low main-sequence stars
concentrated around $V-I \simeq 0.5$ mag. The spread of normal MS
stars in $R-{\rm H}\alpha$ is mostly due
to the photometric errors in the narrow-band H\alp\ magnitudes. The
group of cooler stars with $R-{\rm H}\alpha \geq 0.4$ mag are mostly
faint PMS stars, which exhibit H\alp\ excess.  Few bright stars with
strong H\alp\ emission, which may be Herbig Ae/Be stars, or Be type stars,
are also found.  {\em Right}: The $V-I$, $V$ CMD of all 
emission line sources (thick symbols) overlayed on the CMD of the whole
sample of stars.  Both bright and faint PMS
stars with H\alp\ excess can be readily identified, but the
nature of the stars located on the main sequence just below the turn-off
is still unclear. The rather bright detection limit for the emission
stars is due to the low sensitivity of the H\alp\ filter.\label{fig-haemit}}
\end{figure*}

\section{H\alp\ Observations of the NGC~346/N~66 Region\label{sec-ha}}

\subsection{Stars with H\alp\ excess}

In total, more than 23,000 stars are detected with our photometry in all
three $V$, $I$ and H\alp\ bands with photometric uncertainties less than
0.1 mag. By comparing the H\alp\ and $R$ magnitudes of the detected
sources, one can identify stars with H\alp\ emission (e.g. Grebel et al.
1992, Keller et al. 1999). As our dataset is lacking observations in
$R$, we determined a ``synthetic''$R$ magnitude for every star derived
by accurate interpolation between $V$-, $R$- and $I$-band data for the
filter set of ACS, as ${\rm m}_{675}=0.38{\rm m}_{555}+0.62{\rm
m}_{814}$ (De Marchi, {\em private communication}). We identified the
H\alp\ emission stars in our sample by plotting the $R-{\rm H}\alpha$
color index against case $V-I$ (see Fig.~\ref{fig-haemit}, {\em left}),
and then applying the selection criterion requiring all H\alp\ emission
stars to have $R-{\rm H}\alpha \geq 0.4$ mag above the sequence of
non-emission stars (Keller et al. 1999). This selection revealed 309
stars, which are marked by fat dots in the $V-I$, $V$ CMD in Fig.\
\ref{fig-haemit} ({\em right}).  The bright sources of this sample are
candidate Herbig Ae/Be stars or evolved Be stars, while the fainter are
T~Tauri stars still in their pre-main sequence phase.

\subsection{Spatial Distribution of Stars with H\alp\ Excess}

The spatial distribution of the stars with H\alp\ excess coincides with
the outline of the H{\sc ii} region (Fig.~\ref{fig-ima} ({\em right}) as
well as the general spatial distribution of the PMS stars (Fig. 
\ref{fig-cms}, {\em right}) and the infrared emission from {\em Spitzer}
observations (Simon et al. 2007; see also next section). Two major
concentrations of emission stars can be readily identified, one located
in the nebula N~66, following its general southeast-northwest trend, and
the other to the north-northeast part of the association, coinciding
with clusters 2a, 2b and 3.

\section{{\em Spitzer} Observations of NGC~346\label{sec-spitzer}}

\begin{figure}[t!]
\epsscale{1.15}
\plotone{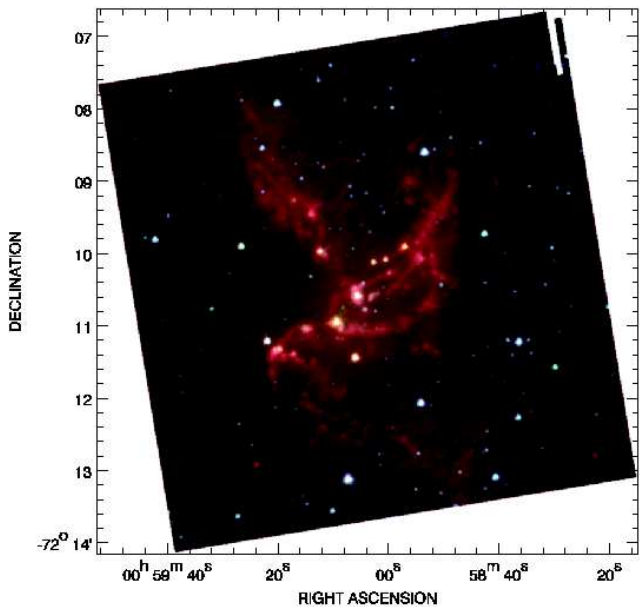}
\caption{Color-composite image of NGC~346/N~66 from {\em Spitzer}/IRAC 
with 3.6 \micron\ shown in blue, 4.5 \micron\ in green, and 8 \micron\ in red.}
\label{fig-mirima}
\end{figure}

As part of the {\em Spitzer} Survey of the Small Magellanic Cloud
(S$^3$MC), Bolatto et al.\ (2007) imaged the SMC in all seven MIPS and
IRAC wavebands, and compiled a photometric catalog of 400,000 mid- and
far-infrared point sources. A color-composite image from {\em
Spitzer}/IRAC imaging of the general area of NGC~346/N~66, retrieved
from the {\em Spitzer} Data Archive is shown in Fig.~\ref{fig-mirima}.
Bolatto et al. (2007) identified 282 bright candidate Young Stellar
Objects (YSOs) as bright 5.8 \micron\ sources with very faint optical
counterparts and very red mid-infrared colors ([5.8]$-$[8.0] $>$ 1.2
mag, where [5.8] and [8.0] are the magnitudes of the sources in the 5.8
\micron\ and 8.0 \micron\ IRAC bands, respectively.  16 of these
candidate YSOs fall within the NGC 346 region observed with HST/ACS.  By
comparison with our ACS photometric catalog (Paper I) we obtained
optical photometry for the candidate YSOs. Because of the difference in
angular resolution of at least $\approx$20 between {\em Spitzer} and
ACS, a unique identification of the optical counterparts of the YSO
candidates is not always possible.  In Table \ref{tab-yso} we summarize
the IRAC and MIPS photometry of the YSOs identified by Bolatto et al.\
(2007) as well as the ACS photometry of the brightest star located
within 1\farcs5 of the {\em Spitzer} position. We also quote the number
of stars detected on ACS within this radius, and find evidence for
significant clustering. Up to $\approx$20 brighter main sequence and
fainter pre-main sequence stars can be found within 1\farcs5 of the
YSOs, i.e.\ we might be seeing compact stellar clusters. The mean
density of these stars for the whole region ($\sim$ 0.15
stars/$\Box$\arcsec) is much lower than the average stellar density of
the stars found within 1\farcs5 around candidate YSOs. A study of the
clustering of PMS stars at the loci of YSOs in the SMC association
NGC~602 is presented by Gouliermis et al. (2007). These authors note
that high-resolution near-infrared imaging and spectroscopy are required
to better understand the stellar populations and clarify if they are
signatures of on-going clustered star formation.

\begin{deluxetable*}{ccccccccrcc}
\tablewidth{0pt}
\tabletypesize{\scriptsize}
\tablecaption{Candidate YSOs found with {\em Spitzer} in the area centered 
on NGC~346 observed with ACS. ID numbers shown in column 1 are from 
Bolatto et al. 2007. Celestial coordinates are given in J2000. Magnitudes 
in all four IRAC channels (3.6 \micron\, 4.5 \micron\, 5.8 \micron\, 8.0 
\micron) and the 24 \micron\ channel of MIPS are given in columns 4, 5, 6, 7 
and 8 respectively. None of these objects was detected in the longer 
wavebands of MIPS (70 \micron\ and 160 \micron). The Numbers of the stars 
found to coincide from our ACS photometry (Paper I) with the loci of 
these YSOs (within 1\farcs5) and the $F555W$ and $F814W$ ($V$, $I$) VEGA 
magnitudes of the brightest ones are given in the last three 
columns.\label{tab-yso}}
\tablehead{
\colhead{ID} &
\colhead{RA (deg)} &
\colhead{DEC (deg)} &
\colhead{[3.6]} &
\colhead{[4.5]} &
\colhead{[5.8]} &
\colhead{[8.0]} &
\colhead{[24]} &
\colhead{No} &
\colhead{$m_{\rm F555W}$} &
\colhead{$m_{\rm F814W}$}
}
\startdata
185	&	14.7150	&	$-$72.1710	&	-	&       14.91	&	12.62	&	10.33	&-& 4&22.55&22.04\\
186	&	14.7167	&	$-$72.1673	&	-	&	14.56	&	12.58	&	11.21	&-& 5&18.39&17.41\\
187	&	14.7190	&	$-$72.1602	&	15.17	&	15.86	&	12.11	&	10.31	&-& 7&23.14&22.50\\
188	&	14.7202	&	$-$72.1783	&	15.75	&	14.97	&	12.33	&	10.64	&-& 4&23.82&21.64\\
189	&	14.7231	&	$-$72.1624	&	14.57	&	15.32	&	12.52	&	10.97	&-& 1&18.86&17.76\\
190	&	14.7275	&	$-$72.1642	&	14.57	&	15.28	&	12.16	&	10.55	&-& 8&20.04&19.46\\
191	&	14.7514	&	$-$72.1681	&	-	&	12.02	&	10.64	&	~~9.19	&5.43& 6&16.27&15.96\\
192	&	14.7582	&	$-$72.1762	&	14.76	&	14.35	&	12.59	&	11.25	&-&13&18.99&19.13\\
194	&	14.7605	&	$-$72.1687	&	12.38	&	11.48	&	10.43	&	~~9.22	&4.41&10&18.67&17.68\\
195	&	14.7622	&	$-$72.1765	&	13.43	&	13.26	&	11.49	&	10.02	&-& 5&14.45&14.70\\
196	&	14.7631	&	$-$72.1774	&	13.93	&	13.73	&	11.78	&	10.24	&-& 9&18.39&18.26\\
197	&	14.7672	&	$-$72.1797	&	-	&	12.74	&	11.67	&	10.31	&-& 8&16.41&15.11\\
198	&	14.7973	&	$-$72.1670	&	-	&	13.91	&	11.90	&	10.02	&-& 7&15.91&16.09\\
200	&	14.8008	&	$-$72.1663	&	11.78	&	11.28	&	~~9.96	&	~~8.43	&5.10& 2&16.77&16.04\\
203	&	14.8208	&	$-$72.1544	&	13.90	&	13.58	&	11.87	&	10.12	&-&19&19.95&19.51\\
204	&	14.8293	&	$-$72.1585	&	-	&	14.83	&	12.67	&	11.12	&-& 9&19.39&18.41
\enddata
\end{deluxetable*}

More recently, Simon et al. (2007) used the data of the S$^{3}$MC survey
and detected a significant population of bright, red infrared sources. 
They expanded the catalog of candidate YSOs in the general area of N~66,
and carried out spectral energy distribution (SED) fits based on models
by Whitney et al. (2003a,b, 2004) to select $\approx$110 YSOs and YSO
candidates. Of these, 65 fall within the field of NGC~346 observed with
HST/ACS. The positions of these candidate YSOs, kindly provided by J.D.\
Simon, are overplotted on the isodensity contour map of the PMS stars in
the region of NGC~346 in Fig.~\ref{fig-cms} (right). The majority of the
candidate YSOs from the Simon et al.\ catalog are located where we find
the highest densities of PMS stars. Specifically, almost all of the
``definite'' YSOs (solid circle symbols) coincide with the density peaks
of the map of Fig.~\ref{fig-cms} (right), and most of them are located
in the association and its surrounding area. In addition, four of the
PMS clusters we identified earlier (\S 4) coincide with ``definite''
YSOs (clusters 2, 3, 4, and 5). According to Simon et al. (2007) the
spatial distribution of the YSOs reveal that star formation in N~66 did
not proceed from the southwest toward the center of the {\sc H ii}
region as proposed by Massey et al (1989). Simon et al. also note that
the most-embedded YSOs are more centrally concentrated than the more
evolved ones, and that the YSOs are mass segregated with the most
massive ones being located closer to the center of N~66. Also, the
clumps of molecular gas found by Rubio et al. (2000) from CO(2-1)
emission, and the peaks of dust emission at 7\micron\ analyzed by
Contursi et al. (2000) coincide with many of these YSOs.

\section{Summary}

In this paper we present the results from our photometric study on the
recent star formation history of N~66 the brightest {\sc H~ii} region in
the SMC, related to the OB association NGC~346, as it is recorded in the
observed PMS population of the region. Our deep photometry revealed an
extremely large number of low-mass PMS stars in the association and the
surrounding region, easily distinguishable in the $V-I$, $V$ CMD. We
show that factors such as reddening, binarity and variability can cause
a broadening of the positions of these stars in the CMD of the
association, wide enough so that they can be misinterpreted as the
result of multi-epoch star formation, and not as the product of a single
star formation event. We found that a modest reddening like the one we
found for the general area of NGC~346 ($E(B-V)\simeq 0.08$ mag) can make
the PMS stars appear younger than what they actually are, and therefore
an age of 10 Myr better fits the observed sequence of PMS stars. 
However, our results do not exclude the possibility of multi-epoch star
formation in the area of the association if the reddening is even lower,
as suggested from the OB stars of this region ($E(B-V)\simeq 0.04$ mag).
In this case, two star formation events 10 and 5 Myr ago can explain the
observed broadening of the PMS stars due to age spread and factors such
as reddening and binarity. No specific dependence of the estimated ages
of the PMS stars to their loci within the association, as a signature of
sequential star formation, was found.

It has been previously suggested that three different generations of
stars occurred through sequential star formation in the region of
NGC~346/N~66 (Rubio et al. 2000) within the last 3 Myr (based on the age
of the youngest OB stars) and that the PMS stars of NGC~346 represent a
star formation event that took place 3 - 5 Myr ago (Nota et al. 2006).
However, the study by Massey et al. (1989) based on the presence of
evolved 15~M{\solar} stars suggests that there might have been earlier
star formation events in the region making NGC~346 as old as $\sim$~12
Myr. In addition, there are recent indications that NGC~346 might host
classical Be-type stars (Evans et al. 2006), and if so, the age of the
association should be at least 10~Myr, a threshold given by Fabregat \&
Torrej{\'o}n (2000) for classical Be-type stars to form. These results
fit very well to the hypothesis that star formation in NGC~346 did
already occur about 10~Myr ago, as our observations and simulations of
the PMS stars suggest.

From star counts based on our ACS photometry we identify at least five
PMS clusters across the region, covering a range of ages. On spatial
scales from 0.8$''$ to 8$''$ (0.24 to 2.4\,pc at the distance of the
SMC) the clustering of the PMS stars as computed by a two-point angular
correlation function is self-similar with a power law slope $\gamma
\approx -0.3$. The clustering properties are quite similar to Milky Way
star-forming regions like the Orion OB association or $\rho$\,Oph. Thus
molecular cloud fragmentation in the SMC seems to proceed on the same
spatial scales as in the Milky Way. This is remarkable given the
differences in metallicity and hence dust content between SMC and Milky
Way star-forming regions.

The youngest PMS stars are located mostly to the north of the bar of
N~66, where three PMS clusters are identified. This area is also
characterized by a high concentration of candidate YSOs (Simon et al.
2007), H\alp-excess stars (found with our photometry), and IR-emission
peaks (Rubio et al.  2000). This indicates that star formation probably
still takes place in an arc-like feature, as it is outlined by the
spatial distribution of these sources. In an accompanying letter, we
combine these results with previous multi-wavelength studies of the
region to investigate the star formation history, which helped to shape
NGC~346/N~66 (Gouliermis et al. 2007b).

\acknowledgments

D. A. Gouliermis kindly acknowledges the support of the German Research
Foundation (Deutsche Forschungsgemeinschaft - DFG) through the
individual grant GO 1659/1-1. This paper is based on observations made
with the NASA/ESA Hubble Space Telescope, obtained from the data archive
at the Space Telescope Science Institute. STScI is operated by the
Association of Universities for Research in Astronomy, Inc. under NASA
contract NAS 5-26555. It is also based on observations made with the
Spitzer Space Telescope, which is operated by the Jet Propulsion
Laboratory, California Institute of Technology under a contract with
NASA.




\begin{references}
\reference{} Blaauw, A. 1983, IrAJ, 16, 141
\reference{} Brice\~{n}o, C., et al. 2001, Sci, 291, 93
\reference{} Brice{\~n}o, C., et al. 2005, AJ, 129, 907
\reference{} Brice\~{n}o, C., et al. 2007, in ``Protostars and Planets V'', 
Eds. B. Reipurth, D. Jewitt, and K. Keil, University of Arizona Press, 
Tucson, p. 345
\reference{} Bolatto, A., et al. 2007,ApJ, 655, 212
\reference{} Chu, Y.-H.\ 1997, \aj, 113, 1815 
\reference{} Chu, Y.-H., \& Kennicutt, R.~C., Jr.\ 1988, \aj, 95, 1111 
\reference{} Contursi, A., et al. 2000, A\&A, 362, 310
\reference{} Danforth, C.~W., Sankrit, R., Blair, W.~P., Howk, J.~C., \&
Chu, Y.-H.\ 2003, \apj, 586, 1179
\reference{} Davies, R.~D., Elliott, K.~H., \& Meaburn, J.\ 1976, MmRAS,
81, 89
\reference{} de Bruijne, J.~H.~J., Hoogerwerf, R., \& de Zeeuw, P.~T.\ 
2001, A\&A, 367, 111
\reference{} Dolphin, A.~E.\ 2000, \pasp, 112, 1383 
\reference{} Duquennoy, A., \& Mayor, M.\ 1991, A\&A, 248, 485 
\reference{} Efremov, Y.~N., \& Elmegreen, B.~G.\ 1998, \mnras, 299, 588 
\reference{} Elmegreen, B. G.\ 1998, PASA, 15, 74
\reference{} Elmegreen,B. G.\ 2006, in ``Globular Clusters, Guide to 
Galaxies'' ed. T. Richtler, et al., ESO/Springer 
({\tt arXiv:astro-ph/0605519})
\reference{} Evans, C.~J., Lennon, D.~J., Smartt, S.~J., \& Trundle, C.\ 
2006, \aap, 456, 623
\reference{} Elson, R.~A.~W., Sigurdsson, S., Davies, M., Hurley, J., \& 
Gilmore, G.\ 1998, MNRAS, 300, 857
\reference{} Fabregat, J., \& Torrej{\'o}n, J.~M.\ 2000, \aap, 357, 451 
\reference{} Fukuda N. \& Hanawa T. 2000, ApJ, 533, 911
\reference{} Girardi, L.,et al. 2002, A\&A, 391, 195
\reference{} Gomez, M., Hartmann, L., Kenyon, S.~J., \& Hewett, R.\
1993, \aj, 105, 1927
\reference{} Goodwin, S. P., Kroupa, P., Goodman, A., Burkert, A. 2007,
in ``Protostars and Planets V'', Eds. B. Reipurth, D. Jewitt, and K.
Keil, University of Arizona Press, Tucson, p. 951
\reference{} Gouliermis, D., Brandner, W., \& Henning, T.\ 2006a, ApJL,
636, L133
\reference{} Gouliermis, D.~A., Dolphin, A.~E., Brandner, W., \&
Henning, T.\ 2006b, ApJS, 166, 549 (Paper I)
\reference{} Gouliermis, D., Brandner, W., \& Henning, T.\ 2005, \apj,
623, 846
\reference{} Gouliermis, D.~A., Quanz, S.~P., \& Henning, T.\ 2007, ApJ, 
665, 306
\reference{} Gouliermis, D., et al.\ 2007b, submitted to ApJL
\reference{} Grebel, E. K., Richtler, T., \& de Boer, K. S. 1992, A\&A, 
254, L5
\reference{} Haberl, F., \& Sasaki, M.\ 2000, \aap, 359, 573 
\reference{} Henize, K.~G.\ 1956, ApJS, 2, 315 
\reference{} Herbst, W., Herbst, D. K., Grossmann, E. J., Weinstein, D. 
1994, AJ, 108, 1906
\reference{} Johnson, H. L., Morgan, W. W. 1953, ApJ, 117, 313
\reference{} Keller, S. C., Wood, P. R., \& Bessell, M. S. 1999, A\&AS, 
134, 489
\reference{} K{\"o}hler, R., Petr-Gotzens, M.~G., McCaughrean, M.~J.,
Bouvier, J., Duch{\^e}ne, G., Quirrenbach, A., \& Zinnecker, H.\ 2006,
A\&A, 458, 461
\reference{}Kontizas, E., Kontizas, M., Gouliermis, D., Dapergolas, A.,
Korakitis, R., Morgan, D. H. 1999, IAUS, 190, 410
\reference{} Larson, R.~B.\ 1995, \mnras, 272, 213 
\reference{} Massey, P., Parker, J. W., Garmany, C. D. 1989, AJ, 98, 1305
\reference{} Mathieu, R.~D.\ 1994, ARA\&A, 32, 465 
\reference{} Nakajima, Y., et al. 1998, ApJ, 497, 721
\reference{} Naz\'{e}, Y., et al. 2002, ApJ, 580, 225
\reference{} Niemela, V. S., Marraco, H. G., \& Cabanne, M. L. 1986, PASP, 
98, 1133
\reference{} Nota, A., et al. 2006, ApJ, 640, L29
\reference{} Palla, F., \& Stahler, S.~W.\ 1999, ApJ, 525, 772
\reference{} Palla, F., \& Stahler, S.~W.\ 2000, ApJ, 540, 255 
\reference{} Petr, M.~G., Coude Du Foresto, V., Beckwith, S.~V.~W.,
Richichi, A., \& McCaughrean, M.~J.\ 1998, ApJ, 500, 825
\reference{} Pollock, A. M. T. 2002, in ASP Conf. Ser., 260, Interacting
Winds from Massive Stars, ed. A. F. J. Moffat \& N. St Louis (San
Francisco: ASP), 363
\reference{} Preibisch, T., \& Zinnecker, H., 1999, AJ, 117, 2381
\reference{} Prosser, C.~F., Stauffer, J.~R., Hartmann, L., Soderblom,
D.~R., Jones, B.~F., Werner, M.~W., \& McCaughrean, M.~J.\ 1994, ApJ,
421, 517
\reference{} Pudritz, R.~E.\ 2002, Science, 295, 68 
\reference{} Rela\~{n}o, M., Peimbert, M., \& Beckmann, J. 2002, ApJ, 564, 
704
\reference{} Rochau, B., Gouliermis, D.~A., Brandner, W., Dolphin,
A.~E., \& Henning, T.\ 2007, ApJ, 664, 322
\reference{} Rubio, M., Contursi, A., Lequeux, J., Probst, R., Barbá, R., 
Boulanger, F., Cesarsky, D., Maoli, R. 2000, A\&A, 359, 1139
\reference{} Sabbi, E., et al. 2007, AJ, 133, 44
\reference{} Sanduleak, N.\ 1968, AJ, 73, 34
\reference{} Scalo, J. M.\ 1985 in Protostars and Planets II, University
of Arizona Press, 201
\reference{} Sherry, W. H.\ 2003, Ph.D.~Thesis, State University of New 
York at Stony Brook
\reference{} Sherry,W. H., Walter, F. M., Wolk, S. J. 2004 AJ, 128, 2316
\reference{} Shu, F.~H., Adams, F.~C., \& Lizano, S.\ 1987, \araa, 25, 23 
\reference{} Siess, L., Dufour, E., \& Forestini, M. 2000, A\&A, 358, 593
\reference{} Simon, J. D., et al.\ 2007, accepted for publication in ApJ
({\tt arXiv:astro-ph/0707.3998})
\reference{} Stahler, S. W., \& Palla, F. 2005, The Formation of Stars,
ISBN 3-527-40559-3, Wiley-VCH
\reference{} Tout, C. A. 1991, MNRAS, 250, 701
\reference{} Vanhala, H. A. T., \& Cameron, A. G. W. 1998, ApJ, 508, 291
\reference{} V\'azquez-Semadeni, E. 2004, Ap\&SS, 292, 187
\reference{} Vel\'{a}zquez, P. F., Koenigsberger, G., \& Raga, A. C.
2003, ApJ, 584, 284
\reference{} Walborn, N. R., et al. 2000, PASP, 112, 1243
\reference{} Walborn, N. R. 1978, APJ, 224, L133
\reference{} Walborn, N. R., \& Blades, J. C. 1986, ApJ, 304, L17
\reference{} Wessolowski, U. 1996, in MPE Rep. 263, R\"{o}ntgenstrahlung
from the Universe, ed. H. U. Zimmerman, J. E. Tr\"{u}mper, \& H. Yorke,
75
\reference{} Whitney, B. A., Wood, K., Bjorkman, J. E., \& Wolff, M. J. 
2003, ApJ, 591, 1049
\reference{} Whitney, B. A., Wood, K., Bjorkman, J. E., \& Cohen, M. 2003,
ApJ, 598, 1079
\reference{} Whitney, B. A., Indebetouw, R., Bjorkman, J. E., \& Wood, K.
2004, ApJ, 617, 1177
\reference{} Woitas, J., Leinert, C., K\"{o}hler, R.\ 2001, A\&A, 376, 982
\reference{} Ye, T., Turtle, A. J., \& Kennicutt, R. C., Jr. 1991,
MNRAS, 249, 722
\end{references}
\end{document}